\documentclass[runningheads]{llncs}
\usepackage[T1]{fontenc}
\usepackage[ruled, linesnumbered, vlined]{algorithm2e}
\usepackage{amsmath}
\usepackage{amssymb}
\usepackage{graphicx}
\usepackage{xcolor}
\usepackage{multirow}
\usepackage{caption}
 
\newcommand{\Tau}{\mathcal{T}}
\begin{document}
\title{Quorum Tree Abstractions of Consensus Protocols}
%
%
\author{Berk Cirisci\inst{1}\orcidID{0000-0003-4261-090X} \and
Constantin Enea\inst{2}\orcidID{0000-0003-2727-8865} \and
Suha Orhun Mutluergil\inst{3}\orcidID{0000-0002-0734-7969}}

\authorrunning{B. Cirisci et al.}
%
\institute{IRIF, Universit\'e Paris Cit\'e \\
\email{cirisci@irif.fr} \and
LIX, Ecole Polytechnique, CNRS and Institut Polytechnique de Paris \\
\email{cenea@lix.polytechnique.fr} \and
Sabanci University \\
\email{suha.mutluergil@sabanciuniv.edu}}
\maketitle              
\begin{abstract}
Distributed algorithms solving agreement problems like consensus or state machine replication are essential components of modern fault-tolerant distributed services. They are also notoriously hard to understand and reason about. Their complexity stems from the different assumptions on the environment they operate with, i.e., process or network link failures, Byzantine failures etc. In this paper, we propose a novel abstract representation of the dynamics of such protocols which focuses on quorums of responses (votes) to a request (proposal) that form during a run of the protocol. We show that focusing on such quorums, a run of 
a protocol can be viewed as working over a tree structure where different branches represent different possible outcomes of the protocol, the goal being to stabilize on the choice of a fixed branch. This abstraction resembles the description of recent protocols used in Blockchain infrastructures, e.g., the protocol supporting Bitcoin or Hotstuff. We show that this abstraction supports reasoning about the safety of various algorithms, e.g., Paxos, PBFT, Raft, and HotStuff, in a uniform way. In general, it provides a novel induction based argument for proving that such protocols are safe.
\end{abstract}
%
%
%

\newcommand{\add}{\mathit{add}}
\newcommand{\commit}{\mathit{commit}}
\section{Introduction}

Consensus or state-machine replication protocols are essential ingredients for maintaining strong consistency in modern fault-tolerant distributed systems.  
Such protocols must execute in the presence of concurrent and asynchronous message exchanges as well as benign (message loss, process crash) or Byzantine failures (message corruption). Developing practical implementations or reasoning about their correctness is notoriously difficult. Standard examples include the classic Paxos~\cite{DBLP:journals/tocs/Lamport98} or PBFT~\cite{DBLP:conf/osdi/CastroL99} protocols, or the more recent HotStuff~\cite{DBLP:conf/podc/YinMRGA19} protocol used in Blockchain infrastructures.


In this paper, we propose a new abstraction for representing the executions of such protocols that can be used in particular, to reason about their safety, i.e., ensuring  \emph{Agreement} (e.g., all correct processes decide on a single value) and \emph{Validity} (e.g., the decided value has been proposed by some node participating in the protocol). Usually, protocol executions are composed of a number of communication-closed rounds~\cite{DBLP:journals/scp/ElradF82}, and each round consists of several phases in which a process broadcasts a request and expects to collect responses from a quorum
of processes before advancing to the next phase. The abstraction is defined as a \emph{sequential} object called \emph{Quorum Tree (QTree)} which maintains a tree structure where each node corresponds to a different round in an execution. The operations of QTree, to add or change the status of a node, model quorums of responses that have been received in certain phases of a round.

For instance, a round in single-decree Paxos consists of two phases: a \emph{prepare} phase where a pre-determined leader broadcasts a request for joining that round and expects a quorum of responses from the other processes before advancing to a \emph{vote} phase where it broadcasts a value to agree upon and expects a quorum of responses (votes) in order to declare that value as decided in that round. Rounds are initiated by their respective leaders and can run concurrently. The idea behind QTree is to represent a Paxos execution using a rooted tree where each node different from the root corresponds to a round where the leader has received \emph{a quorum of responses in the prepare phase}. The parent-child relation models the data flow from one round to a later round: responses to join requests contain values voted for in previous rounds (if any) and one of them will be included by the leader in the vote phase request. The round in which that value was voted defines the parent. Then, each node has one out of three possible statuses: \texttt{ADDED} if the vote phase can still be successful (the leader can collect a quorum of votes) but this did not happen yet, \texttt{GHOST} if the vote phase can not be successful (e.g., a majority of processes advanced to the next round without voting), and \texttt{COMMITTED} if the leader has received a quorum of responses in the vote phase. This is a tree structure because before reaching a quorum in the vote phase of a round, other rounds can start and their respective leaders can send other vote requests (with possibly different values).
 The specific construction of requests and responses in Paxos ensures that all the \texttt{COMMITTED} nodes in this tree belong to a \emph{single} branch, which entails the agreement property (this will become clearer when presenting the precise definition of QTree in Section~\ref{sec:qtree}).

The QTree abstraction is applicable to a wide range of protocols beyond the single-decree Paxos sketched above. It applies to state-machine replication protocols like Raft~\cite{DBLP:conf/cpp/WoosWATEA16} and HotStuff~\cite{DBLP:conf/podc/YinMRGA19} where the tree structure represents logs of commands (inputted by clients) stored at different processes and organized according to common prefixes (each node corresponds to a single command) and multi-decree consensus protocols like multi-Paxos~\cite{DBLP:journals/tocs/Lamport98} and its variants~\cite{DBLP:journals/corr/cs-DC-0408036,malkhi2008stoppable,DBLP:journals/dc/Lamport06,DBLP:journals/corr/HowardMS16}, or PBFT~\cite{DBLP:conf/osdi/CastroL99} where different consensus instances (for different indices in a sequence of commands) are modeled using different QTree instances. 

We show that all these protocols are \emph{refinements} of QTree in the sense that their executions can be mapped to sequences of operations on a QTree state, which are about agreeing on a branch of the tree called the \emph{trunk}. These operations are defined as invocations of two methods $\add$ and $\commit$ for adding a new leaf to the tree (during which some other nodes may turn to \texttt{GHOST}) and changing the status of a node from \texttt{ADDED} to \texttt{COMMITTED}, respectively. Any sequence of invocations to these methods ensures that all the \texttt{COMMITTED} nodes lie on the same branch of the tree (the trunk). In relation to protocol executions, $\add$ and $\commit$ invocations that concern the same node correspond to receiving a quorum of responses in two specific phases of a round, which vary from one protocol to another. 

The mapping between protocol executions and QTree executions is defined as in proofs of linearizability for concurrent objects with \emph{fixed} linearization points. Analogous to linearizability, where the goal is to show that an object method takes effect instantaneously at a point in time called linearization point, we show that it is possible to mark certain steps of a given protocol as linearization points of $\add$ or $\commit$ operations\footnote{These linearization points are \emph{fixed} in the sense that they correspond to specific instructions in the code of the protocol, and they do not depend on the future of an execution. For an expert reader, this actually corresponds to a proof of strong linearizability~\cite{DBLP:conf/stoc/GolabHW11}.}, such that the sequence of $\add$ and $\commit$ invocations defined by the order between linearization points along a protocol execution is a correct QTree execution.
We introduce a declarative characterization of correct QTree executions that simplifies the proof of the latter (see Section~\ref{ssec:trans_sys}).

The QTree abstraction offers a novel view on the dynamics of classic consensus or state-machine replication protocols like Paxos, Raft, and PBFT, which relates to the description of recent Blockchain protocols like HotStuff and Bitcoin~\cite{bitcoin}, i.e., agreeing on a branch in a tree. It provides a formal framework to reason \emph{uniformly} about single-decree consensus protocols and state-machine replication protocols like Raft and HotStuff. For single-decree protocols (or compositions thereof), the parent-child relation between QTree nodes corresponds to the data-flow between a quorum of responses to a leader and the request he sends in the next phase while for Raft and HotStuff, it corresponds to an order set by a leader between different commands.

Our work relies on a hypothesis that correctness proofs based on establishing a refinement towards an \emph{operational} specification such as QTree, which can be understood as a sequence of steps, are 
much more intuitive and ``explainable'' compared to classic proofs based on inductive invariants. 
An inductive invariant has to describe all intermediate states produced by all possible orders of receiving messages and a precise formalization is quite complex. As an indication, the Paxos invariant used in recent work~\cite{DBLP:journals/pacmpl/PadonLSS17} (see formulas (4) to (12) in Section 5.2) is a conjunction of eight quantified first-order formulas which are hard to reason about and not re-usable in the context of a different protocol. 

We believe that operational specifications are also helpful in taming complexity while designing new protocols or implementations theoreof, or in gaining confidence about their correctness without going through ad-hoc and brittle proof arguments. For instance, our proofs are very clear about the phases of a round in which quorums need to intersect, which provides flexibility and optimization opportunities for deciding on quorum sizes in each phase. Depending on environment assumptions, quorum sizes can be optimized while preserving correctness.
Compared to previous operational specifications for reasoning about consensus protocols, e.g.,~\cite{DBLP:journals/sigact/BoichatDFG03,DBLP:conf/esop/Garcia-PerezGMS18}, QTree is designed to be less abstract so that the refinement proof, establishing the relationship between a given protocol and QTree, is less complex (see Section~\ref{sec:related} for details). 


%
%
%

\section{Quorum Tree}\label{sec:qtree}
We describe the QTree sequential object which operates on a tree and has two 
methods $\add$ and $\commit$ for adding a new node and modifying an attribute of a node (committing a node), respectively. When used as an abstraction of consensus protocols, invocations of these two methods correspond to certain quorums that are reached during a round of the protocol.
\subsection{Overview}


QTree is a sequential rooted-tree, a possible state being depicted in Figure~\ref{fig:qtree}. The nodes with black dashed margins are not members of the tree and they are discussed later. 
Each node in the tree contains a round number, a value, and a status field set to \texttt{ADDED}, \texttt{GHOST}, or \texttt{COMMITTED}. The round number acts as an identifier of a node since there can not exist two nodes with the same round number.
The $Root$ node is part of the initial state and its status is \texttt{COMMITTED}.
A QTree state consists of a trunk, alive branches, and dead branches; a branch is a chain of nodes connected by the parent relation. Alive branches are extensible with new \texttt{ADDED} nodes but dead branches are not.
The trunk is a particular branch of the tree that starts from the root. It contains all the \texttt{COMMITTED} nodes and it ends with a \texttt{COMMITTED} node. It may also contain \texttt{ADDED} or \texttt{GHOST} nodes. 
For example, in Figure~\ref{fig:qtree}, the trunk consists of $Root$ and $n_3$. 
All alive branches are connected to the last \texttt{COMMITTED} node of the trunk (alive branches can include \texttt{ADDED} or \texttt{GHOST} nodes). For instance, in Figure~\ref{fig:qtree}, the subtree rooted at $n_3$ contains a single alive branch whose leaf node is $n_5$. Dead branches can contain only \texttt{GHOST} nodes. In Figure~\ref{fig:qtree}, the tree contains a single dead branch containing the node $n_1$.

Nodes can be added to the tree as leaves. The status of a newly added node is either \texttt{ADDED} or \texttt{GHOST}. The status \texttt{ADDED} may turn to \texttt{GHOST} or \texttt{COMMITTED}. 
The \texttt{GHOST} status is ``final'' meaning that it can never turn into \texttt{COMMITTED} afterwards. However, \texttt{GHOST} nodes can be part of alive branches, and they can help in growing the tree. 

QTree has two methods $\add$ and $\commit$: 
\begin{itemize}
\vspace{-1mm}
	\item $\add$ generates a new leaf with a round number $r$ value $v$ and parent $p$ identified by the round number $r_p$ given as an input. Its status is set to \texttt{ADDED} or \texttt{GHOST} provided that some conditions hold. If the status of the new node is set as \texttt{ADDED}, then it either extends (has a path to the end of) an existing alive branch or creates a new alive branch from the trunk. The new node may also ``invalidate'' some other nodes by changing their status from \texttt{ADDED} to \texttt{GHOST}. 
	\item $\commit$ extends the trunk by turning the status of a node from \texttt{ADDED} to \texttt{COMMITTED}. This extension of the trunk may prevent some branches to be extended in the future (some alive branches may become dead), i.e., future invocations of $\add$ that extend those branches will add only \texttt{GHOST} nodes.
\end{itemize}
\vspace{-.5mm}
Each node models the evolution of a round in a consensus protocol and the value attribute represents the value proposed by the leader of that round. The round and value attributes of a node are immutable and cannot be changed later. We assume that round numbers are strictly positive except for $Root$ whose round number is 0. 

QTree applies uniformly to a range of consensus or state-machine replication protocols. We start by describing a variation that applies to single-decree consensus protocols, where a number of processes aim to agree on a single value. Multi-decree consensus protocols that are used to solve state-machine replication can be simulated using a number of instances of QTree, one for each decree (the instances are independent one from another). Then, state-machine replication protocols like HotStuff that rely directly on a tree structure to order commands can be simulated by the QTree for single-decree consensus modulo a small change that we discuss later.

%
%
\vspace{-1.5mm}
\subsection{Definition of the Single-Decree Version}

\begin{figure}[t]
\begin{minipage}[p]{.63\textwidth}
\begin{algorithm}[H]
\SetAlCapNameFnt{\scshape}
\caption{The QTree object}
\footnotesize
\label{alg:general}
\SetKwProg{init}{Initialize: \mbox{\quad /* $\bot$ denotes non-initialized values */}}{}{end}
\SetKwProg{propose}{\emph{Method Propose}}{}{end}
\SetKwProg{add}{Method \emph{add}}{}{end}
\SetKwProg{commit}{Method \emph{commit}}{}{end}
\SetKwIF{If}{ElseIf}{Else}{if}{}{else if}{else}{}
\SetKwFor{ForAll}{forall}{}{}

\init{}{$Root$.round = \textbf{0};
	$Root$.status = \texttt{COMMITTED}\;
	$Root$.value = $\bot$;
	$Root$.parent = \textbf{$Root$}\;
	Nodes = $\{Root\}$;
}

\add{($r$, $v$, $r_p$)}{\label{alg:joinStart}
     \textbf{Pre:} $r$ > 0 \\
    	$n$ = \textbf{new} Node(round = $r$, status = $\bot$,
	value = $v$, parent = $p: p.\mathit{round} = r_p$)\;\label{alg:nondet}
         \If{$valid(n)$ $\land$ {\color{red} \underline{$valueConstraint(n) $}}}{\label{alg:isValid}
        	      Nodes = Nodes $\cup$ \{$n$\}\;\label{alg:add_node}
	      $n$.status = \texttt{ADDED}\;\label{alg:added} 
              \If{$\exists n' \in Nodes.\ n'$.round $> n$.round}{ \label{line:if}
              	 $n$.status = \texttt{GHOST}\; 
              } 
              \ForAll{$n' \in Nodes.\ n'$.round $< n$.round}{ \label{line:if_ghost}
       	     	\If{$n$ is conflicting with n$'$}{
              		$n'$.status $\gets$ \texttt{GHOST}\;
         	   }
    	       }
	      \Return OK
      }
      \Return FAIL
}\label{alg:joinEnd}

\commit{($r$)}{
     \If{$\exists$ $n$ $\in$ Nodes. $n$.round = $r$ $\land$ \linebreak $n$.status = \emph{\texttt{ADDED}} }{\label{alg:star2} 
          $n$.status $\gets$ \texttt{COMMITTED}\; 
          \Return OK
      }
      \Return FAIL
}
\end{algorithm}
\end{minipage}\hfill
\begin{minipage}[tp]{.37\textwidth}
\centering
\includegraphics[width=1\linewidth]{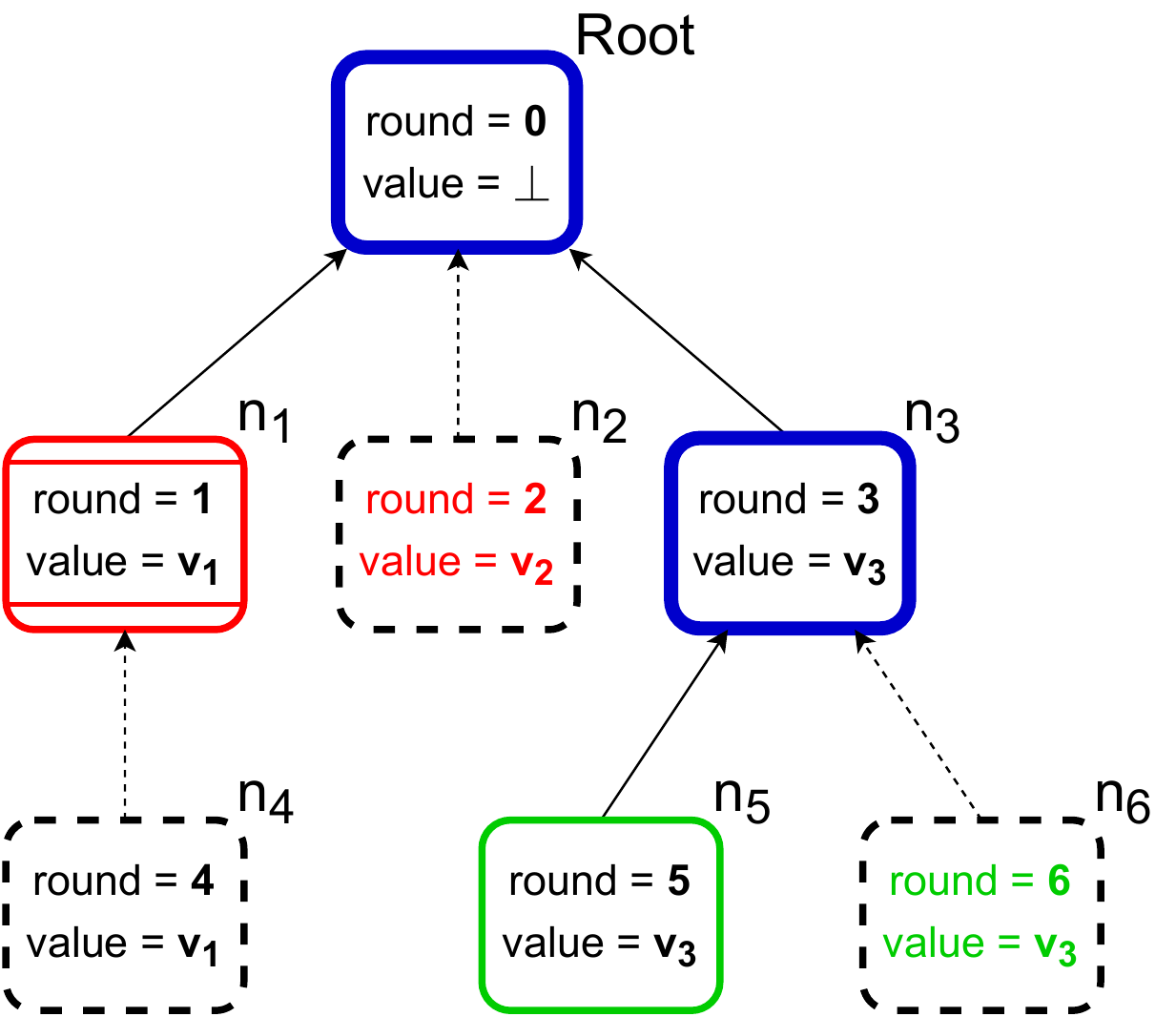}
\vspace*{-6mm}
\captionof{figure}{A state of QTree. We represent \texttt{ADDED} nodes with green solid margins, \texttt{GHOST} nodes with red double-line margins, and \texttt{COMMITTED} nodes with blue thick margins. The nodes with black dashed margins are not part of the state, they are fictitious nodes used to explain the method for adding new nodes.} 
\label{fig:qtree}
\end{minipage}
\vspace{-6mm}
\end{figure}

Algorithm~\ref{alg:general} lists a description of QTree in pseudo-code. 
The following set of predicates are used as conditions inside methods:
\vspace{-1mm}
\begin{enumerate}
\item	$link(n)$ $\equiv$ $n$.parent $\in$ Nodes $\land$ $n$.parent.round < $n$.round\; \label{pred:link}
\item	$newRound(n) \equiv \forall n' \in$ Nodes. $n'$.round $\neq$ $n$.round\; \label{pred:newRound}
\item	$maxCommitted(n) \equiv n$.status = \texttt{COMMITTED} $\land$ \\($\forall n' \in$ Nodes. $n'$.status = \texttt{COMMITTED} $\implies n'$.round < $n$.round)\;
\item	$extendsTrunk(n) \equiv \exists n' \in$ Nodes. $maxCommitted(n')$ $\land$ \\($n$ extends $n' \lor n$.round $< n'$.round)\; \label{pred:extendsTrunk}
\item	$valid(n) \equiv link(n) \land newRound(n) \land extendsTrunk(n)$\; \label{pred:valid}
\item	$valueConstraint(n) \equiv n$.parent $\neq Root \implies n$.value $=$ $n$.parent.value \label{pred:valueConst}
\end{enumerate} 
The $\add$ method (lines \ref{alg:joinStart}-\ref{alg:joinEnd}) generates a new node $n$ with round, value, and parent set according to the method's inputs. 
Then, it adds $n$ to the tree by linking it to the selected parent if $n$ satisfies the following \emph{validity} conditions:
%
%
\vspace{-2mm}
\begin{itemize}
	\item $n$'s parent belongs to the tree and its round number is smaller than $r$ (predicate \emph{link} at~(\ref{pred:link})),
	\item the tree does not contain a node with round number $r$ (predicate \emph{newRound} at~(\ref{pred:newRound})),
	\item if $r$ is bigger than the round number of the last node of the trunk, then $n$ must extend the trunk (predicate \emph{extendsTrunk} at~(\ref{pred:extendsTrunk})),
	\item $n$'s value must be the same as its parent's value unless the parent is the $Root$ (predicate \emph{valueConstraint} at~(\ref{pred:valueConst})).
\vspace{-2mm}
\end{itemize}
The \emph{valid} predicate at~(\ref{pred:valid}) is the conjunction of the first three constraints.

For example, let us consider an invocation of $\add$ in a state of QTree that contains the non-dashed nodes in Figure~\ref{fig:qtree}. If the invocation generates $n_2$, $n_4$, or $n_6$ (receiving as input the corresponding attributes), then $n_2$ and $n_6$ do satisfy all these constraints and can be added to the tree. The node $n_4$ fails the \emph{extendsTrunk} predicate because it is not extending the last node of the trunk ($n_3$) and its round number is higher. 




If a node $n$ satisfies the conditions above, the $\add$ method turns its status to either \texttt{ADDED} or \texttt{GHOST}. If there is another node in the tree with a higher round number, $n$'s status becomes \texttt{GHOST}. Otherwise, it becomes \texttt{ADDED}. As a continuation of the example above, the status of $n_2$ is set to \texttt{GHOST} because the tree contains node $n_3$ with a higher round number and the status of $n_6$ is set to \texttt{ADDED}.

\begin{figure}[t]
\centering
	\includegraphics[width=1\linewidth]{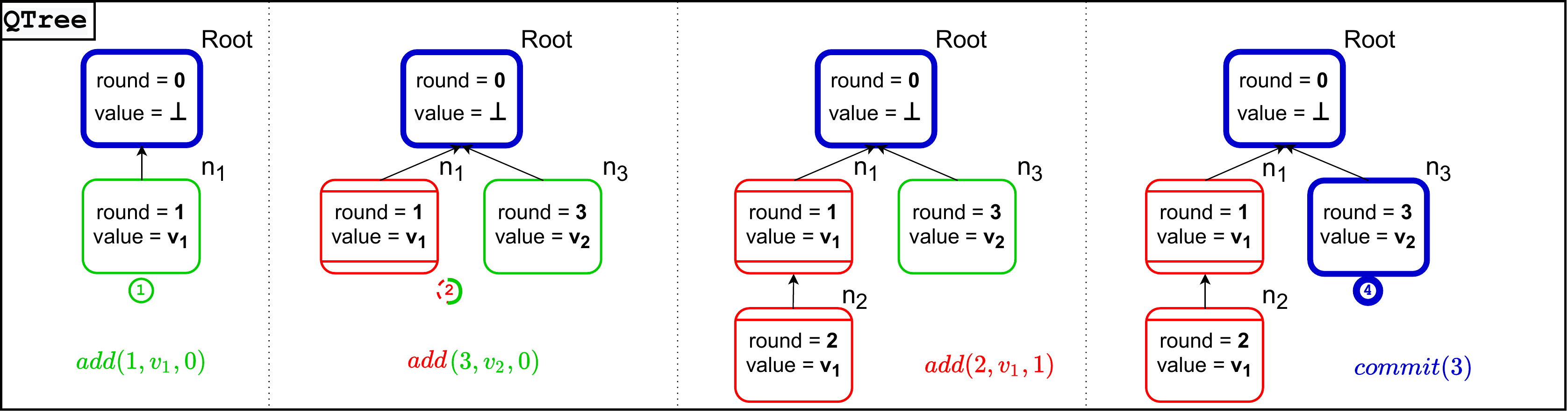}
	\vspace{-5mm}
	\caption{Explaining the behavior of $\add$ and $\commit$ methods. Colors are interpreted as in Fig~\ref{fig:qtree}.}
	\label{fig:all}
\vspace{-5mm}
\end{figure}

Moreover, the addition of $n$ can ``invalidate'' some other nodes, turn their status to \texttt{GHOST}. This is based on a notion of \emph{conflicting} nodes. We say that two nodes are conflicting if they are on different branches, i.e., there is no path from one node to the other. An $\add$ invocation that adds a node $n$ changes 
 the status of all the nodes $n'$ in the tree that conflict with $n$ and have a lower round number than $n$, to \texttt{GHOST}. For example, Figure~\ref{fig:all} pictures a sequence of QTree states in an execution, to be read from left to right. The first state represents the result of executing $\add(1,v_1,0)$ on the initial state of QTree, adding node $n_1$. Executing $\add(3,v_2,0)$ on this first state creates another node $n_3$ and sets its status to \texttt{ADDED}. This invocation will also turn the status of $n_1$ to \texttt{GHOST} since its round number is less than the round number of $n_3$ and they are on different branches. Afterwards, by executing $\add(2,v_1,1)$, a node $n_2$ is added to the tree with status \texttt{GHOST} since there is a node $n_3$ on a different branch which has a higher round number.

The method $\add$ returns $OK$ when the created node is effectively added to the tree (it satisfies the conditions described above) and $FAIL$, otherwise.


Lastly, the $\commit$ method takes a round number $r$ as input and turns the status of the node containing $r$ to \texttt{COMMITTED} if it was \texttt{ADDED}. If successful, it returns $OK$ and $FAIL$, otherwise. As a continuation of the example above, the right part of Figure~\ref{fig:all} pictures a state obtained by executing $\commit(3)$ on the state to the left. This sets the status of $n_3$ to \texttt{COMMITTED} as $n_3$ was previously \texttt{ADDED}.
Note that the conditions in $\add$ ensure that the tree can not contain two nodes with the same round number.

\medskip
\noindent
\textbf{Safety Properties.} 
We show that the QTree object in Algorithm~\ref{alg:general} can be used to reason about the safety of single-decree consensus protocols, in the sense that it satisfies a notion of \emph{Validity} (processes agree on one of the proposed values) and \emph{Agreement} (processes decide on a single value). 
More precisely, we show that every state that is reachable by executing a sequence of invocations of $\add$ and $\commit$ (in Algorithm~\ref{alg:general}), called simply \emph{reachable state}, satisfies the following:
\vspace{-1mm}
\begin{itemize}
\item \emph{Validity}: every node different from $Root$ contains the same value as a child of $Root$, and
\item \emph{Agreement}: every two $\texttt{COMMITTED}$ nodes different from $Root$ contain the same value.
\vspace{-1mm}
\end{itemize}


\vspace{-1mm}
\begin{proposition}[Validity]\label{the:validity}
Every node in a reachable state that is different from $Root$ contains the same value as a child of $Root$.
\vspace{-1mm}
\end{proposition}
\begin{proof}
A node $n$ is added to the tree only if the predicate \emph{valueConstraint} holds, which implies that it is either a child of $Root$ or it has the same value as its parent which is a descendant of $Root$. Also, since the value attribute of a node is immutable, any $\texttt{COMMITTED}$ node contains the same value that it had when it was created by an $\add$ invocation. 
\end{proof}

\vspace{-1.5mm}
Therefore, the fact that a consensus protocol refining QTree satisfies validity, i.e., processes decide on a value proposed by a client of the protocol, reduces to proving that the phases of a round simulated by $\add$ invocations that add children of $Root$ use values proposed by a client. This is ensured using additional mechanisms, i.e., a client broadcasts its value to all participants in the protocol, so that each participant can check the validity of a value proposed by a leader. 


%
%
%

Next, we focus on \emph{Agreement}, and show that \texttt{COMMITTED} nodes belong to a single branch of the tree.

\vspace{-1mm}
\begin{proposition}
	\label{the:CommittedChain}
	Let $n_1$ and $n_2$ be two \texttt{COMMITTED} nodes in a reachable state. Then, $n_1$ and $n_2$ are not conflicting.
\vspace{-1mm}
\end{proposition}
\begin{proof}
	Assume towards contradiction that QTree reaches a state where two \texttt{COMMITTED} nodes $n_1$ and $n_2$ are conflicting. Let $r_1=n_1\text{.round}$ and $r_2=n_2\text{.round}$. Without loss of generality, we assume that $r_1<r_2$. Such a state is reachable if $\add(r_1,\_,\_)$ and $\add(r_2,\_,\_)$ resulted in adding the nodes $n_1$ and $n_2$ and set their status to \texttt{ADDED} (we use $\_$ to denote arbitrary values), and subsequently, $\commit(r_1)$ and $\commit(r_2)$ switched the status of both $n_1$ and $n_2$ to \texttt{COMMITTED}.
	 If $\add(r_1,\_,\_)$  were to execute before $\add(r_2,\_,\_)$, then $\add(r_2,\_,\_)$  would have changed the status of $n_1$ to \texttt{GHOST} because it is conflicting with $n_2$. Otherwise, if $\add(r_2,\_,\_)$  were to execute before $\add(r_1,\_,\_)$ , then the latter would have set the status of $n_1$ to \texttt{GHOST} since the tree contains $n_2$ that has a higher round number. In both cases, executing $\commit(r_1)$ can never turn the status of $n_1$ to \texttt{COMMITTED}. 	
\end{proof}

\vspace{-1.5mm}
Proposition~\ref{the:CommittedChain} allows to conclude that any two \texttt{COMMITTED} nodes (different from $Root$) contain the same value. Indeed, a node can become \texttt{COMMITTED} only if it was \texttt{ADDED}, which implies that is has the same value as its parent (the predicate \emph{valueConstraint} holds), and by transitivity, as any of its ancestors, except for $Root$.

\vspace{-2mm}
\begin{proposition}[Agreement]
	\label{the:CommittedValue}
	Let $n_1$ and $n_2$ be two \texttt{COMMITTED} nodes in a reachable state, which are different from $Root$. Then, $n_1$.value = $n_2$.value.
\vspace{-2mm}
\end{proposition}

%
%
%
%
%
%
\vspace{-1.5mm}
\subsection{State Machine Replication Versions}

The single-decree version described above can be extended easily to a \emph{multi-decree} context. As multi-decree consensus protocols, used in state machine replication, can be seen as a composition of multiple instances of single-decree consensus protocols, a multi-decree version of QTree is obtained by composing multiple instances of the single-decree version. Each of these instances manipulates a tree as described above without interference from other instances. The validity and agreement properties above apply separately to each instance.

%

The single-decree version can also be extended for state machine replication protocols like HotStuff and Raft where the commands (values) are a-priori structured as a tree, i.e., each command given as input is associated to a predetermined parent in this tree. Then, the goal of such a protocol is to agree on a sequence in which to execute these commands, i.e., a branch in this tree. Simply removing the \emph{valueConstraint} condition in the $\add$ method (underlined in Algorithm~\ref{alg:general})  enables QTree to simulate such protocols. A node's value need not be the same as its parent's value to be valid for $\add$. Proposition~\ref{the:CommittedChain} that implies the agreement property of such protocols still holds (Proposition~\ref{the:CommittedValue} does not hold when the \emph{valueConstraint} condition is removed; this property is specific to single-decree consensus). Since the value field remains immutable, the validity property of such protocols reduces to ensuring that the values generated during phases simulated by $\add$ correspond to commands issued by the client (Proposition~\ref{the:validity} is also specific to single-decree consensus and it does not hold). As before, this requires additional mechanisms, i.e., a client broadcasting a command to all the participants in the protocol, whose correctness can be established quite easily.

\vspace{-1.5mm}
\section{Consensus Protocols Refining QTree}\label{ssec:trans_sys}

In the following, we show that a number of consensus protocols are refinements of QTree in the sense that their executions can be mimicked with $\add$ and $\commit$ invocations. This is similar to a linearizable concurrent object being mimicked with invocations of a sequential specification.
The refinement relation allows to conclude that the \emph{Validity} and \emph{Agreement} properties of QTree imply similar properties for any of its refinements.

The definition of the refinement relation relies on a formalization of protocols and QTree as \emph{labeled transition systems}. For a given protocol, a state is a tuple of process local states and a set of messages in transit, and a transition corresponds to an indivisible step of a process (receiving a set of messages, performing a local computation step, or sending a message). For QTree, a state is a tree of nodes as described above and a step corresponds to an invocation to $\add$ or $\commit$. An execution is a sequence of transitions from the initial state.

Refinement corresponds to a mapping between protocol executions and QTree executions. 
This mapping is defined as in proofs of linearizability for concurrent objects with \emph{fixed linearization points}, where the goal is to show that each concurrent object method appears to take effect instantaneously at a point in time that corresponds to executing a fixed statement in its code. 
Therefore, certain steps of a given protocol are considered as linearization points of $\add$ and $\commit$ QTree invocations (returning $OK$), and one needs to prove that the sequence of invocations defined by the order of linearization points in a protocol execution is a correct execution of QTree.

Formally, a labeled transition system (LTS) is a tuple $L = (\mathcal{Q}, q_0, \mathcal{T}, \mathcal{A}_L)$ where  $\mathcal{Q}$ is a set of states, $q_0$ is the unique initial state, $\mathcal{A}_L$ is a set of actions (transition labels) and $\Tau$ is a set of transitions $(q,a,q')$ such that $q,q' \in \mathcal{Q}$ and $a \in \mathcal{A}_L$. An \emph{execution} $E$ from $q_0$ is a finite sequence of alternating states and actions such that $E=q_0,a_0,q_1,a_1,\ldots,q_n$ with $(q_i,a_i,q_{i+1})\in\Tau$ for each $0\leq i\leq n-1$. A trace $t$ is the sequence of actions projected from some execution $E$.
$T(L)$ denotes the set of traces of $L$. 

The standard notion of refinement between LTSs states that an LTS $L$ is a refinement of another LTS $L'$ when $T(L) \subseteq T(L')$.
In this paper, we consider a slight variation of this definition of refinement that applies to LTSs that do \emph{not} share the same set of actions, representing for instance, some concrete protocol and QTree, respectively. This notion of refinement is parametrized by a mapping $\Gamma$ between actions of $L$ and $L’$, respectively. We say that \emph{$L$ $\Gamma$-refines $L'$} when $\Gamma(T(L)) \subseteq T(L')$. Here, a mapping $\Gamma : \mathcal{A}_L \rightarrow \mathcal{A}_{L’}$ is extended to sequences and sets of sequences as expected, e.g., $\Gamma(a_1 \ldots a_n) = \Gamma(a_1) \ldots \Gamma(a_n)$. With this extension, the preservation of safety specifications from an LTS to a refinement of it requires certain constraints on the mapping $\Gamma$ that will be discussed in Section~\ref{ssec:safetyQt}. 

In the context of proving that a concrete protocol refines QTree, the goal is to define a mapping $\Gamma$ between actions of the protocol and QTree $\add/\commit$ invocations such that $\Gamma$ applied to protocol executions results in correct QTree executions. In the following, we provide a characterization of correct QTree executions that simplifies such refinement proofs. 
\subsection{Characterizing QTree Invocation Sequences}
\label{sec:invSeq}

An \emph{invocation label} $\add(r,v,r_p)\Rightarrow RET$ or $\commit(r)\Rightarrow RET$ combines a QTree method name with input values and a return value $RET\in \{OK, FAIL\}$. An invocation label is called \emph{successful} when the return value is $OK$. A sequence $\sigma$ of invocation labels is called \emph{correct} when there exist QTree states $q_0$, $\ldots$, $q_{|\sigma|}$, such that $q_0$ is the QTree initial state and for each $i\in [1,|\sigma|]$, executing $\sigma_i$ starting from $q_{i-1}$ leads to $q_i$.\\

\begin{theorem}\label{th:main}
A sequence $\sigma$ of successful invocation labels is correct if and only if the following hold (we use $\_$ to denote arbitrary values):
\begin{enumerate}
\item\label{eq:prop1} for every $r$, $\sigma$ contains at most one invocation label $\add(r,\_,\_)$ and at most one invocation label $\commit(r)$
\item\label{eq:prop0} every $\commit(r)$ is preceded by an $\add(r,\_,\_)$
\item\label{eq:prop2} if $r_p > 0$, every $\add(r,v,r_p)$ is preceded by $\add(r_p,v',\_)$ where $0 < r_p < r$
	\begin{enumerate}
	\item \label{eq:prop2a} and $\mathit{v = v'}$
	\end{enumerate}
\item\label{eq:prop3} if $\sigma$ contains $\add(r,\_,\_)$ and $\add(r',\_,r'')$ with $r''<r<r'$, then $\sigma$ does \emph{not} contain $\commit(r)$
\end{enumerate}
\end{theorem}
Properties~\ref{eq:prop1}--\ref{eq:prop2} are straightforward consequences of the $\add$ and $\commit$ definitions. Indeed, it is impossible to add two nodes with the same round number $r$, which implies that there can not be two successful $\add(r,\_,\_)$ invocations, the status of a node can be flipped to $\texttt{COMMITTED}$ exactly once, which implies that there can not be two successful $\commit(r)$ invocations, and a $\commit(r)$ is successful only if a node with round number $r$ already exists, hence Property~\ref{eq:prop0} must hold. Moreover, a node's parent defined by the input $r_p$ must already exist in the tree, which implies that Property~\ref{eq:prop2} must also hold. Property~\ref{eq:prop3} is more involved and relies on the fact that a node $n$ with round number $r$ can be $\texttt{COMMITTED}$ only if there exist no other conflicting node $n'$ with a bigger round number $r'$ (the parent of $n'$ having a round smaller than $r$ implies that $n$ and $n'$ are conflicting). 

\begin{proof}
\noindent
($\Rightarrow$): 
Assume that $\sigma$ is correct. We show that it satisfies the above properties:
\begin{itemize} 
\item Property \ref{eq:prop1}: The $newRound(n)$ predicate used at line \ref{alg:isValid} in Algorithm~\ref{alg:general} ensures that it is impossible to add two nodes with the same round number $r$, and therefore $\sigma$ can not contain two successful $\add(r,\_,\_)\Rightarrow OK$ invocations. The conditions at  line \ref{alg:star2} ensure that $\commit(r)\Rightarrow OK$ can flip the status of a node only once, and therefore only one such successful invocation can occur in $\sigma$.
\item Property \ref{eq:prop0}: The conditions at line \ref{alg:star2} in Algorithm~\ref{alg:general} imply that the state in which $\commit(r)\Rightarrow OK$ is executed contains a node with round number $r$. This node could have only added by a previous $\add(r,\_,\_)\Rightarrow OK$ invocation.
\item Property \ref{eq:prop2}: The $link(n)$ predicate used at line  \ref{alg:isValid} in Algorithm~\ref{alg:general} ensures that the state in which $\add(r,v,r_p)\Rightarrow OK$ is executed contains a node with round number $r_p$. This node could have only added by a previous $\add(r_p,v',\_)\Rightarrow OK$ invocation, for some $v'$. 
	\begin{itemize}
	\item Property \ref{eq:prop2a}: It is a direct consequence of the $valueConstraint(n)$ predicate used at line  \ref{alg:isValid} in Algorithm~\ref{alg:general}.
	\end{itemize}
\item Property \ref{eq:prop3}: Let $n$ and $n'$ be the nodes of the QTree state $q$ reached after executing $\sigma$, which have been added by $\add(r,\_,\_)\Rightarrow OK$ and $\add(r',\_,r'')\Rightarrow OK$, respectively. We have that $n'$.round $> n$.round $> n$.parent.round. Since the round numbers decrease when going from one node towards $Root$ in a reachable QTree state, it must be the case that $n$ and $n'$ are conflicting. By Lemma \ref{lem:invariant}, we get that $n.\text{status}$ is \texttt{GHOST}. Since the \texttt{GHOST} status can not be turned to \texttt{COMMITTED}  and vice-versa, it follows that $\sigma$ can not contain $\commit(r)\Rightarrow OK$.
\end{itemize}

\medskip
\noindent
($\Leftarrow$): We prove that every sequence $\sigma$ that satisfies properties \ref{eq:prop1}--\ref{eq:prop3} is correct.
 We proceed by induction on the size of $\sigma$. The base step is trivial. For the induction step, let $\sigma$ be a sequence of size $k+1$. If $\sigma$ satisfies properties \ref{eq:prop1}-\ref{eq:prop3}, then the prefix $\sigma'$ containing the first $k$ labels of $\sigma$ satisfies properties \ref{eq:prop1}-\ref{eq:prop3} as well. By the induction hypothesis, $\sigma'$ is correct. We show that the last invocation of $\sigma$, denoted by $\sigma_{k+1}$ can be executed in the QTree state $q_{|\sigma'|}$ reached after executing $\sigma'$. We start with a lemma stating an inductive invariant for reachable QTree states:

\begin{lemma}
For every node $n$ in any state $q$ reached after executing a correct sequence $\sigma$ of successful invocations, $n.\text{status}$ is \texttt{COMMITTED} if $n$ is $Root$ or $\sigma$ contains a $\commit(r)$ invocation. Else, $n.\text{status}$ is \texttt{GHOST} if $q$ contains a node $n'$ with $n'.\text{round} > n.\text{round}$ and $n'$ is conflicting with $n$, and it is \texttt{ADDED}, otherwise.
\label{lem:invariant}
\end{lemma}
\begin{proof}
We proceed by induction on the size of $\sigma$. The base step is trivial. For the induction step, let $\sigma$ be a sequence of size $m+1$. Let $q_m$ be the state reached after executing the prefix of size $m$ of $\sigma$, and let $\sigma_{m+1}$ be the last invocation label of $\sigma$. We show that the property holds for any possible $\sigma_{m+1}$ that takes the QTree state $q_{m}$ to some other state $q_{m+1}$:
\begin{itemize}
\item $\sigma_{m+1}= \add(r,v,r_p)\Rightarrow OK$, for some $r$, $v$, $r_p$: Let $n$ be the new node added by this invocation. 
The status of $n$ can be \texttt{ADDED} or \texttt{GHOST}. If $q_m$ contains a node $n'$ with $n'.\text{round} > r$ (since round numbers are decreasing going towards the $Root$ and $n$ is a new leaf node, any existing node with a higher round number such as $n'$ is also conflicting with $n$), then the status of $n$ becomes \texttt{GHOST} by the predicate at line \ref{line:if} in Algorithm~\ref{alg:general} (otherwise, it remains \texttt{ADDED}). This implies that $n$'s status satisfies the statement in the lemma. This invocation may also turn the status of some set of nodes $N$ from \texttt{ADDED} to \texttt{GHOST} by the statement at line \ref{line:if_ghost} in Algorithm~\ref{alg:general}. 
The nodes in $N$ 
have a lower round number than $r$ and conflicting with $n$. Therefore, the statement of the lemma is satisfied for the nodes in $N$.
\item $\sigma_{m+1}=\commit(r)\Rightarrow OK$, for some $r$: For $\commit(r)$ to be successful the conditions at line \ref{alg:star2} in Algorithm~\ref{alg:general} must be satisfied. If it is satisfied, only the status of node $n$ is changed from \texttt{ADDED} to \texttt{COMMITTED}. Note that $Root$ exists by definition and its status is \texttt{COMMITTED}. Since the statuses of the rest of the nodes stay the same, the statement of the lemma holds.\hfill$\Box$
\end{itemize}
\end{proof}

 There are two cases to consider depending on whether $\sigma_{k+1}$ is an $\add$ or $\commit$ invocation label:
\begin{itemize}
\item $\add(r,v,r_p)$:  This invocation label is successful if and only if the predicates $valid(n)$ and $valueConstraint(n)$ at line \ref{alg:isValid} in Algorithm~\ref{alg:general} are satisfied after generating a new node $n$ with the given inputs in the state $q_{|\sigma'|}$:
	\begin{itemize}
	\item $newRound(n)$: Due to Property \ref{eq:prop1}, $r \neq n'$.round for any other node $n' \in q_{|\sigma'|}$ and the predicate is satisfied.
	\item $link(n)$: To satisfy this predicate, there must exist a node in $q_{|\sigma'|}$ with round $r_p$  where $r_p < r$.  By Property \ref{eq:prop2}, if $\sigma$ contains $\add(r,\_,r_p)\Rightarrow OK$ with $r_p \neq 0$, then $\add(r_p,\_,\_)\Rightarrow OK$ also exists in $\sigma$. Hence, there exists a node $p$ with round $r_p$ in $q_{|\sigma'|}$, and the predicate is satisfied. If $r_p = 0$, then $q_{|\sigma'|}$ contains the $Root$ node (with round 0) which ensures that the predicate is satisfied. 
	\item $extendsTrunk(n)$: This predicate states that $n$ extends the node $n'$ which has the highest round number among the nodes with \texttt{COMMITTED} status, if $n$.round > $n'$.round. Assume by contradiction that this is not the case, i.e., $n$.round > $n'$.round but $n$ and $n'$ are conflicting. Let $n_1$ be the lowest common ancestor of $n$ and $n'$ (the first common node on the paths from $n$ and $n'$ to the $Root$). Since the round numbers decrease when going from one node towards $Root$, we have that $n_1$.round < $n'$.round. If we consider the nodes on the path from $n$ to $n_1$, since $n$.round > $n'$.round, there must exist a node $n_2$ such that $n_2$.round > $n'$.round but $n_2$.parent.round < $n'$.round. The node $n_2$ in $q_{|\sigma'|}$ corresponds to the invocation label $\add(n_2$.round,$\_,n_2$.parent.round$)$ in $\sigma'$. Moreover, the \texttt{COMMITTED} status of $n'$ implies the existence of $\commit(n'$.round$)$ in $\sigma'$ as stated in Lemma \ref{lem:invariant}. However, it is impossible that $\sigma'$ contains both these invocation labels if Property~\ref{eq:prop3} holds. 
	\item $valueConstraint(n)$: It is implied trivially as Property \ref{eq:prop2a} holds.
	\end{itemize}
\item $\commit(r)$: It is successful if and only if the conditions at line \ref{alg:star2} in Algorithm~\ref{alg:general} are satisfied. Then by Property \ref{eq:prop1} and \ref{eq:prop0}, there exist $\add(r,\_,\_)$ in $\sigma'$ but not $\commit(r)$. As $\add(r,\_,\_)$ is successful, there already exist a node $n$ in $q_{|\sigma'|}$ where its round is $r$ but its status can be either \texttt{ADDED} or \texttt{GHOST}. Towards a contradiction, assume that $n$.status $=$ \texttt{GHOST} in $q_{|\sigma'|}$. This means that there exists a node $n'$ conflicting with $n$ such that $n'$.round > $n$.round as stated in Lemma \ref{lem:invariant}. Let $n_1$ be the least common ancestor of $n$ and $n'$. Since round numbers are decreasing going towards the $Root$, $n_1$.round < $n$.round. If we consider nodes on the path from $n'$ to $n_1$, there exists a node $n_2$ such that $n_2$.round > $n$.round and $n_2$.parent.round < $n$.round. That's why, there is an invocation label $\add(n_2.round,\_,n_2.parent.round)$ in $\sigma'$. However, $\sigma$ cannot contain both of these invocation labels together according to Property~\ref{eq:prop3}. \hfill$\Box$
\end{itemize}
\end{proof}

\section{Linearization Points}\label{sec:history_events}


We describe an instrumentation of consensus protocols with linearization points of successful QTree invocations, and illustrate it using Paxos as a running example. Section~\ref{sec:hot} and Section~\ref{sec:pbft} will discuss other protocols like HotStuff, Raft, PBFT, and multi-Paxos. This instrumentation defines the mapping $\Gamma$ between actions of a protocol and QTree, respectively, such that the protocol is a $\Gamma$-refinement of QTree. We also discuss the properties of this instrumentation which imply that establishing $\Gamma$-refinement is an effective proof for the safety of the protocol.




The identification of linearization points relies on the fact that protocol executions pass through a number of rounds, and each round goes through several phases (rounds can run asynchronously -- processes need not be in the same round at the same time). 
The protocol imposes a total order over the phases inside a round and among distinct rounds. Processes executing the protocol can only move forward following the total order on phases/rounds. Going from one phase to the next phase in the same round is possible if a quorum of processes send a particular type of message. The refinement proofs require identifying two quorums for each round where a value is first proposed to be agreed upon and then decided. They correspond to linearization points of successful $\add(r,\_,\_)$ and $\commit(r)$, respectively.
%
The linearization point of $\add(r,v,r_p)\Rightarrow OK$ occurs when intuitively, the value $v$ is proposed as a value to agree upon in round $r$. For the protocols we consider, $v$ is determined by a designated leader after receiving a set of messages from a quorum of processes. For single-decree consensus, members of the quorum send the latest round number and value they adopted (voted) in the past and the leader picks a value corresponding to the maximum round number $r_p$. If no one in the quorum has adopted any value yet, then the leader is free to propose any value received from a client, and $r_p$ equals a default value $0$. For state-machine replication protocols like Raft or HotStuff, the round $r_p$ is defined in a different manner -- see Section~\ref{sec:hot} and Appendix~\ref{appendix:raft}. The linearization point of $\commit(r)\Rightarrow OK$ occurs when a  quorum of nodes adopt (vote for) a value $v$ proposed at round $r$. 

By Theorem~\ref{th:main}, proving that the order between linearization points along a protocol execution defines a correct QTree execution reduces to showing Properties~\ref{eq:prop1}--\ref{eq:prop3}. In general, 
Properties~\ref{eq:prop1}--\ref{eq:prop2} are quite straightforward to establish and follow from the control-flow of a process. 
Property~\ref{eq:prop2a} is specific to single-decree consensus protocols or compositions thereof, e.g., (multi-)Paxos and PBFT. It will not hold for Raft or Hotstuff. 
Property~\ref{eq:prop3} is related to the fact that any two quorums of processes intersect in a correct process.

Above, we have considered the case of a protocol that is a refinement of a single instance of QTree. State machine replication protocols that are composed of multiple independent consensus instances, e.g., PBFT (see Section~\ref{sec:pbft}), are refinements of a set of QTree instances (identified using a sequence number) and every linearization point needs to be associated with a certain QTree instance.

\subsection{Linearization Points for Paxos}

For concreteness, we exemplify the instrumentation with linearization points on the single-decree Paxos protocol. We start with a brief description of this protocol that focuses on details relevant to this instrumentation.
%
%
%

Paxos proceeds in rounds and each round has a unique leader. Since the set of processes running the protocol is fixed and known by every process, the leader of each round can be determined by an a-priorly fixed deterministic procedure (e.g., the leader is defined as $r\ mod\ N$ where $r$ is the round number and $N$ the number of processes). 
For each round, the leader acts as a proposer of a value to agree upon.

A round contains two phases. 
In the first phase, the leader broadcasts a {\fontfamily{qcr}\selectfont START} message to all the processes to start the round, executing the  {\bf \fontfamily{qcr}\selectfont START} action below, and processes acknowledge with a {\fontfamily{qcr}\selectfont JOIN} message if some conditions are met, executing the {\bf {\fontfamily{qcr}\selectfont JOIN}} action:
\vspace{-1mm}
\begin{description}
\item[$\bullet$ {\fontfamily{qcr}\selectfont START} Action:] The leader $p$ of round $r > 0$ (the proposer) broadcasts a {\fontfamily{qcr}\selectfont START}($r$) message to all processes.
  
\item[$\bullet$ {\fontfamily{qcr}\selectfont JOIN} Action:] When a process $p'$ receives a {\fontfamily{qcr}\selectfont START}($r$) message, if $p'$ has not sent a {\fontfamily{qcr}\selectfont JOIN} or {\fontfamily{qcr}\selectfont VOTE} message (explained below) for a higher round in the past\footnote{Each process has a local variable $\mathit{maxJoinedRound}$ that stores the maximal round it has joined or voted for in the past and checks whether $\mathit{maxJoinedRound} < r$}, it replies by sending a {\fontfamily{qcr}\selectfont JOIN}($r$) message to the proposer. This message includes the maximum round number ($\mathit{maxVotedRound}$) for which $p'$ has sent a {\fontfamily{qcr}\selectfont VOTE} message in the past and the value ($\mathit{maxVotedValue}$) proposed in that round. If it has not voted yet, these fields are $0$ and $\bot$.
\vspace{-1mm}
\end{description}

If the leader receives {\fontfamily{qcr}\selectfont JOIN} messages from a quorum of processes, i.e., at least $f+1$ processes from a total number of $2f+1$, the second phase starts. The leader broadcasts a {\fontfamily{qcr}\selectfont PROPOSE} message with a value, executing the {\bf \fontfamily{qcr}\selectfont PROPOSE} action below. 
Processes may acknowledge with a {\fontfamily{qcr}\selectfont VOTE} message if some conditions are met, executing a {\bf \fontfamily{qcr}\selectfont VOTE} action. If the leader receives {\fontfamily{qcr}\selectfont VOTE} messages from a quorum of processes, then the proposed value becomes decided (and sent to the client) by executing a {\bf \fontfamily{qcr}\selectfont DECIDE} action:
\vspace{-1mm}
\begin{description}
\item[$\bullet$ {\fontfamily{qcr}\selectfont PROPOSE} Action:] When the proposer $p$ receives {\fontfamily{qcr}\selectfont JOIN}($r$) messages from a quorum of ($f + 1$) processes, it selects the one with the highest vote round number and proposes its value by broadcasting a {\fontfamily{qcr}\selectfont PROPOSE}($r$) message (which includes that value). If there is no such highest round (all vote rounds are 0), then the proposer selects the proposed value randomly simulating a value given by the client (whose modeling we omit for simplicity). 
 
\item[$\bullet$ {\fontfamily{qcr}\selectfont VOTE} Action:] When a process $p'$ receives a {\fontfamily{qcr}\selectfont PROPOSE}($r$) message, if $p'$ has not sent a {\fontfamily{qcr}\selectfont JOIN} or {\fontfamily{qcr}\selectfont VOTE} message for a higher round in the past, it replies by sending a {\fontfamily{qcr}\selectfont VOTE}($r$) message to the proposer with round number $r$. 

\item[$\bullet$ {\fontfamily{qcr}\selectfont DECIDE} Action:] When the proposer $p$ receives {\fontfamily{qcr}\selectfont VOTE}($r$)  messages from a quorum of processes, it updates a local variable called $\mathit{decidedVal}$ to be the value it has proposed in this round $r$. This assignment means that the value is decided and sent to the client.
 \vspace{-1mm}
\end{description}

\noindent
\textbf{Linearization points in Paxos.}
We instrument Paxos with linearization points as follows:
\begin{itemize}
\item the linearization point of $\add(r,v,r')\Rightarrow OK$ occurs when the proposer broadcasts the {\fontfamily{qcr}\selectfont PROPOSE}($r$) message containing value $v$ after receiving a quorum of {\fontfamily{qcr}\selectfont JOIN}($r$) messages (during the {\bf \fontfamily{qcr}\selectfont PROPOSE} action in round $r$). The round $r'$ is extracted from the {\fontfamily{qcr}\selectfont JOIN}($r$) message selected by the proposer. 
\item the linearization point of $\commit(r)\Rightarrow OK$ occurs when the leader of round $r$ updates $\mathit{decidedVal}$ after receiving a quorum of {\fontfamily{qcr}\selectfont VOTE}($r$)  messages (during the {\bf \fontfamily{qcr}\selectfont DECIDE} Action).
\end{itemize}
We illustrate the definition of linearization points for Paxos in relation to QTree executions in Appendix~\ref{app:paxos}.


%

\begin{theorem}
Paxos refines QTree.
\end{theorem}
\begin{proof}
We show that the sequence of successful $\add$ and $\commit$ invocations defined by linearization points along a Paxos execution satisfies the properties in Theorem~\ref{th:main} and therefore, it represents a correct QTree execution:
\vspace{-1mm}
\begin{itemize}
\item \textbf{Property~\ref{eq:prop1}:} Each round has a unique leader and the leader follows the rules of the protocol (no Byzantine failures), thereby, making a single proposal. Therefore, the linearization point of an $\add(r,\_,\_)\Rightarrow OK$ will occur at most once for a round $r$. Since a single value can be proposed in a round, and all processes follow the rules of the protocol, they can only vote for that single value. Thus, at most one linearization point of $\commit(r)\Rightarrow OK$ can occur for a round $r$.
\item \textbf{Property~\ref{eq:prop0}:} This holds trivially as all the processes follow the rules of the protocol and they need to receive a {\fontfamily{qcr}\selectfont PROPOSE}($r$) message (which can occur only after the linearization point of an $\add(r,\_,\_)\Rightarrow OK$) from the leader of round $r$ to send a {\fontfamily{qcr}\selectfont VOTE}($r$) message.
\item \textbf{Property~\ref{eq:prop2}:} By the definition of the {\bf \fontfamily{qcr}\selectfont PROPOSE} action, the proposer selects a highest vote round number $r'$ from a quorum of {\fontfamily{qcr}\selectfont JOIN}($r$) messages that it receives, before broadcasting a {\fontfamily{qcr}\selectfont PROPOSE}($r$) message. If such a highest vote round number $r' > 0$ exists, then there must be a {\fontfamily{qcr}\selectfont VOTE}($r'$) message which is a reply to a {\fontfamily{qcr}\selectfont PROPOSE}($r'$) message. Thus, if the linearization point of $\add(r,\_,r')\Rightarrow OK$ occurs where $r' \neq 0$, then it is preceded by $\add(r',\_,\_)$. Also, by the definition of {\bf \fontfamily{qcr}\selectfont JOIN}, a process can not send a  {\fontfamily{qcr}\selectfont JOIN}($r$) message after a {\fontfamily{qcr}\selectfont VOTE}($r'$) message if $r \ngtr r'$. 
	\begin{itemize}
        \item \textbf{Property~\ref{eq:prop2a}:} By the definition of {\bf
        \fontfamily{qcr}\selectfont PROPOSE}, the proposer selects the {\fontfamily{qcr}\selectfont
        JOIN} message with the highest vote round number and proposes its value. Thus, if the linearization points of both $\add(r,v,r')\Rightarrow OK$ and $\add(r',v',\_)\Rightarrow OK$ occur, then $\mathit{v = v'}$. 
        \end{itemize} 
\item \textbf{Property~\ref{eq:prop3}:} Assume by contradiction that the linearization point of $\commit$\\$(r) \Rightarrow OK$ occurs along with the linearization points of $\add(r,\_,\_)\Rightarrow OK$ and $\add(r',\_,r'')\Rightarrow OK$, for some $r''<r<r'$. The linearization point of $\commit(r)$ occurs because of a quorum of {\fontfamily{qcr}\selectfont VOTE}($r$) messages sent by a set of processes $P_1$, and $\add(r',\_,r'')$ occurs because of a quorum of {\fontfamily{qcr}\selectfont JOIN}($r'$) messages sent by a set of processes $P_2$. Since $P_1$ and $P_2$ must have a non-empty intersection, by the definition of {\bf {\fontfamily{qcr}\selectfont JOIN}}, it must be the case that $r''\geq r$, which contradicts the hypothesis. 
\vspace{-2mm}
\end{itemize}
\end{proof}

The proof of Property~\ref{eq:prop3} relies exclusively on the quorum of processes in the first phase of a round intersecting the quorum of processes in the second phase of a round. It is not needed that quorums in first, resp., second, phases of different rounds intersect. This observation is at the basis of an optimization that applies to non-Byzantine protocols like Flexible Paxos~\cite{DBLP:journals/corr/HowardMS16} or Raft (see Appendix~\ref{appendix:raft} and Appendix~\ref{ssec:MultiPaxos}).

\subsection{Inferring Safety}\label{ssec:safetyQt}


The main idea behind these linearization points is that successful $\add$ and $\commit$ invocations correspond to some process doing a step that witnesses for the receipt a quorum of messages sent in a certain phase of a round. Intuitively, linearization points of successful $\mathit{add}$ invocation occur when some process in some round is certain that a quorum of processes received or will receive the same proposal (same value, parent etc.) for the same round and acts accordingly (sends a message). Such proposal on a value $v$ in a round $r$ is denoted by the linearization point of successful $\mathit{add}(r, v, r')$ for some $r'$.
 On the other hand, the linearization point of a successful $\mathit{commit}(r)$ invocation occurs when a process decides on a value in round $r$ (e.g., after receiving a quorum of votes). Formally, if we denote the actions of a protocol that correspond to linearization points of successful $\mathit{add}(r, v, r')$ and $\mathit{commit}(r)$ invocations using $a_a$ and $a_c$, respectively, then $\Gamma(a_a) = \mathit{add}(r, v, r')\Rightarrow OK$ and $\Gamma(a_c) = \mathit{commit}(r)\Rightarrow OK$. 
 
When the protocol is such a $\Gamma$-refinement of QTree, then, it satisfies agreement and validity. 
If a decision on a value $v$ in a round $r$ of a protocol is the linearization point of a successful $\mathit{commit(r)}$, then by Theorem~\ref{th:main}, the corresponding QTree state contains a node $n$ with $n$.round = $r$, $n$.value = $v$, 
and $n$.status = \texttt{COMMITTED}. For single-decree consensus, Proposition~\ref{the:CommittedValue} ensures that all rounds decide on the same value. For state machine replication protocols like Raft and HotStuff, where the goal is to agree on a sequence of commands, Proposition~\ref{the:CommittedChain} ensures that all the decided values lie on the same branch of the tree which ensures that all processes agree on the same sequence of commands. 

For validity, when $valueConstraint(n)$ is considered, successful $\mathit{add}(r, v, 0)$ invocations represent proposals of client values. Theorem~\ref{th:main} ensures that these invocations correspond to nodes $n$ that are immediate children of $Root$ and for any such node $n$, $n$.value = $v$. Therefore, by Proposition~\ref{the:validity}, we can conclude that only client values can be decided. When $valueConstraint(n)$ is not considered, the fact that the value of each node is obtained from a client is ensured using additional mechanisms that are straightforward, e.g., a client broadcasting a command to all the participants in the protocol.

\section{HotStuff Refines QTree}\label{sec:hot}

We present an instrumentation of HotStuff with linearization points of successful $\add$ and $\commit$ invocations. We use HotStuff as an example of a state machine replication protocol where processes agree over a sequence of commands to execute, and any new command proposed by a leader to the other processes comes with a well-identified immediate predecessor in this sequence. Agreement over a command entails agreement over all its predecessors in the sequence.
This is different from protocols such as multi-Paxos or PBFT, discussed in the next section, where commands are associated to indices in the sequence and they can be agreed upon in any order.
 Appendix~\ref{appendix:raft} presents an instrumentation of Raft which behaves in a similar manner.

In HotStuff, $f$ out of a total of $N = 3f+1$ processes might be Byzantine in the sense that they might show arbitrary behavior and send corrupt or spurious messages. However, they are limited by cryptographic protocols. HotStuff requires that messages are signed using public-key cryptography, which implies that Byzantine processes cannot imitate messages of correct (non-faulty) processes. Additionally, after receiving a quorum of messages, leaders must include certificates in their own messages to prove that a quorum has been reached. These certificates are constructed using threshold signature schemes and correct processes will not accept any message from the leader if it is not certified. Because of Byzantine processes, HotStuff requires quorums of size of $2f+1$ which ensures that the intersection of any two quorums contains at least one correct process.

Each process stores a tree of commands. When a node in this tree (representing some command) is decided, all the ancestors of this node in the tree (nodes on the same branch) are also decided. For a node to become decided, a leader must receive a quorum of messages in 3 consecutive phases after the proposal. After each quorum is established, the leader broadcasts a different certificate to state which quorum has been achieved and the processes update different local variables accordingly, with the same node (if the certificate is valid). These local variables are $preNode$, $votedNode$ and $decidedNode$ in the order of quorums. 

To start a new round, processes send their $preNode$'s to the leader of the next round in {\fontfamily{qcr}\selectfont ROUND-CHANGE(r)} messages and increment their round number.
 After getting a quorum of messages and selecting the $preNode$ with the highest round, 
 the leader broadcasts a {\fontfamily{qcr}\selectfont PROPOSE(r)} message with a new node (value is taken from the client) whose parent is the selected $preNode$. When the message is received by a process, it first checks if the new node extends the selected $preNode$. Then it accepts the new node if the node extends its own $votedNode$ (it is a descendant of $votedNode$ in the tree) or it has a higher round number than the round number of its $votedNode$, and sends\footnote{For all received messages, a correct process also checks if the round number of the node sent by the leader is equal to the current round number of its own, and can send only one message for each phase in each round.} a {\fontfamily{qcr}\selectfont JOIN(r)} message with the same content. 
 In the second (resp., third) phase, if a quorum of {\fontfamily{qcr}\selectfont JOIN(r)} (resp.,  {\fontfamily{qcr}\selectfont PRECOMMIT\_VOTE(r)}) messages is received by the leader, it broadcasts a {\fontfamily{qcr}\selectfont PRECOMMIT(r)} (resp., {\fontfamily{qcr}\selectfont COMMIT(r)}) message, 
 and processes update their $preNode$ (resp., $votedNode$) with the new node, sending a {\fontfamily{qcr}\selectfont PRECOMMIT\_VOTE(r)} (resp., {\fontfamily{qcr}\selectfont COMMIT\_VOTE(r)}) message. 
 In the fourth phase, when the leader receives a quorum of {\fontfamily{qcr}\selectfont COMMIT\_VOTE(r)}, it broadcasts a {\fontfamily{qcr}\selectfont DECIDE(r)} message and processes update their $decidedNode$ accordingly. 
 See Appendix~\ref{ssec:HotStuffdescript} for more details.

For HotStuff, the linearization points of $\add$ and $\commit$ occur with the broadcasts of {\fontfamily{qcr}\selectfont PRECOMMIT}($r$) and {\fontfamily{qcr}\selectfont DECIDE}($r$) messages, respectively, that are \emph{valid} , i.e., (1) they contain certificates for quorums of {\fontfamily{qcr}\selectfont JOIN(r)} or {\fontfamily{qcr}\selectfont COMMIT\_VOTE(r)} messages, respectively, which respect the threshold signature scheme, and (2) they contain the same node as in those messages. 
More precisely,
\begin{itemize}
\item the linearization point of $\add(r,v,r')\Rightarrow OK$ occurs the \emph{first} time when a \emph{valid} {\fontfamily{qcr}\selectfont PRECOMMIT}($r$) message containing node $v$ is sent.
 $r'$ is the round of the node which is the parent of $v$ and it is contained in a previous {\fontfamily{qcr}\selectfont PROPOSE(r)} message
 ($r'$ can be 0 in which case parent of $v$ is a distinguished root node that exists in the initial state).
 
\item the linearization point of $\commit(r)\Rightarrow OK$ occurs the \emph{first} time when a \emph{valid} {\fontfamily{qcr}\selectfont DECIDE}($r$) message is sent.
\vspace{-1mm}
\end{itemize}
Note that a Byzantine leader can send multiple \emph{valid} {\fontfamily{qcr}\selectfont PRECOMMIT}($r$) messages that include certificates for different quorums of {\fontfamily{qcr}\selectfont JOIN(r)} messages. A linearization point occurs when the first such message is sent. Even if processes reply to another valid {\fontfamily{qcr}\selectfont PRECOMMIT}($r$) message sent later, this later {\fontfamily{qcr}\selectfont PRECOMMIT}($r$) message contains the same $preNode$ value, and their reply will have the same content. The same holds for {\fontfamily{qcr}\selectfont DECIDE}($r$) messages. This remark along with the restriction to valid messages and the fact that any two quorums intersect in at least one correct process implies that the sequence of successful $\add$ and $\commit$ invocations defined by these linearization
points satisfies the properties in Theorem~\ref{th:main} and therefore,

\begin{theorem}\label{th:hot}
HotStuff refines QTree.
\end{theorem}

A detailed proof of the theorem above is given in Appendix~\ref{ssec:HotStuffproofProps}.


\section{PBFT Refines QTree}\label{sec:pbft}

The protocols discussed above are refinements of a \emph{single} instance of QTree. State-machine replication protocols based Multi-decree consensus like Multi-Paxos or PBFT can be seen as compositions of a number of single-decree consensus instances that run concurrently, one for each index in a sequence of commands to agree upon, and they are refinements of a set of independent QTree instances. We describe the instrumentation of PBFT and delegate multi-Paxos (and variants) to Appendix~\ref{ssec:MultiPaxos}.

PBFT is a multi-decree consensus protocol in which processes aim to agree on a sequence of values. As in HotStuff, $f$ out of a total number of $3f+1$ processes might be Byzantine and quorums are of size at least $2f+1$. To ensure authentication, messages are signed using public-key cryptography. Messages sent after receiving a quorum of messages in a previous phase include that set of messages as a certificate.


A new round $r$ starts with the leader receiving a quorum of {\fontfamily{qcr}\selectfont ROUND-CHANGE}($r$) messages (like in HotStuff). Each such message from a process $p$ includes the {\fontfamily{qcr}\selectfont VOTE} message with the highest round (similarly to the {\bf \fontfamily{qcr}\selectfont JOIN} action of Paxos) that $p$ sent in the past, for each sequence number that is not yet agreed by a quorum. For an arbitrary set of sequence numbers $sn$, the leader selects the {\fontfamily{qcr}\selectfont VOTE} message with the highest round and broadcasts a {\fontfamily{qcr}\selectfont PROPOSE}($r$,$sn$) message that includes the same value as in the  {\fontfamily{qcr}\selectfont VOTE} message or a value received from a client if there is no such highest round. As mentioned above, this message also includes the {\fontfamily{qcr}\selectfont VOTE} messages that the leader received as a certificate for the selection. When a process receives a {\fontfamily{qcr}\selectfont PROPOSE}($r$,$sn$) message, if $r$ equals its current round, the process did not already acknowledge a {\fontfamily{qcr}\selectfont PROPOSE}($r$,$sn$) message, and the value proposed in this message is selected correctly w.r.t. the certificate, then it broadcasts a {\fontfamily{qcr}\selectfont JOIN}($r$,$sn$) message with the same content (this is sent to all processes not just the leader). If a quorum of {\fontfamily{qcr}\selectfont JOIN}($r$,$sn$) messages is received by a process, then it broadcasts a {\fontfamily{qcr}\selectfont VOTE}($r$,$sn$) message with the same content. If a process receives a quorum of {\fontfamily{qcr}\selectfont VOTE}($r$,$sn$) messages, then the value in this message is decided for $sn$. When a process sends its highest round number {\fontfamily{qcr}\selectfont VOTE} messages to the leader of the next round (in {\fontfamily{qcr}\selectfont ROUND-CHANGE} messages), it also includes the quorum of  {\fontfamily{qcr}\selectfont JOIN} messages that it received before sending the {\fontfamily{qcr}\selectfont VOTE}, as a certificate.

PBFT is a refinement of a set of independent QTree instances, one instance for each sequence number. The linearization points will refer to a specific instance identified using a sequence number, e.g., $sn.\add(r,v,r')$ denotes an $\add(r,v,r')$ invocation on the QTree instance $sn$. Therefore, 
\vspace{-1mm}
\begin{itemize}
\item the linearization point of $sn.\add(r, v, r')\Rightarrow OK$ occurs the \emph{first} time when a process $p$ sends a {\fontfamily{qcr}\selectfont VOTE}($r$, $sn$) message, assuming that $p$ is \emph{honest}, i.e., it already received a quorum of {\fontfamily{qcr}\selectfont JOIN}($r$, $sn$) messages with the same content.
$v$ is the value of the {\fontfamily{qcr}\selectfont VOTE}($r'$, $sn$) message that is included in the {\fontfamily{qcr}\selectfont PROPOSE}($r$,$sn$) message (it is possible that $r'=0$ and $v$ is selected randomly).
\item the linearization point of $sn.\commit(r)\Rightarrow OK$ occurs the \emph{first} time when a process $p$ decides a value for $sn$, assuming that $p$ is \emph{honest}, i.e., it already received a quorum of {\fontfamily{qcr}\selectfont JOIN}($r$, $sn$), resp., {\fontfamily{qcr}\selectfont VOTE}($r$, $sn$), messages with the same content.
\vspace{-1mm}
\end{itemize}
A protocol refines a set of QTree instances identified using sequence numbers when it satisfies Properties~\ref{eq:prop1}-\ref{eq:prop3} in Theorem~\ref{th:main} for each sequence number, e.g., Property~\ref{eq:prop1} becomes for every $sn$ and every $r$, a protocol execution contains a linearization point for at most one invocation $sn.\add(r, \_, \_)\Rightarrow OK$ and at most one invocation  $sn.\commit(r)\Rightarrow OK$. A detailed proof of the following theorem is given in Appendix~\ref{ssec:PBFTproofProps}.

\begin{theorem}\label{th:pbft}
PBFT refines a composition of independent QTree instances.
\end{theorem}
\section{Discussion}

Protocols considered in this work can be grouped under three classes: single-decree consensus (Paxos), multi-decree consensus (PBFT, Multi-Paxos) and state machine replication (Raft, HotStuff)\footnote{This is a slight abuse of terminology since multi-decree consensus protocols are typically used to implement state machine replication.}. We show that they all refine QTree: a single instance for Paxos and HotStuff, and a set of independent instances (one for each sequence number in a command log) for PBFT, Multi-Paxos, and Raft. The more creative parts of the refinement proofs are the identification of $\add$ and $\commit$ linearization points and establishing Property~\ref{eq:prop3} in Theorem~\ref{th:main} which follows from the intersection of quorums achieved in different phases of a round. The other 3 properties in Theorem~\ref{th:main} which guarantee that the linearization points are correct are established in a rather straightforward manner, based on the control-flow of a process participating to the protocol.

The linearization points of successful $\add$ and $\commit$ invocations correspond to some process doing a step that witnesses for the receipt a quorum of messages sent in a certain phase of a round, e.g., the leader broadcasting a {\fontfamily{qcr}\selectfont PROPOSE}($r$) message in Paxos entails that a quorum of {\fontfamily{qcr}\selectfont JOIN}($r$) messages have been sent in the first phase and received. Protocols vary in the total number of phases in a round, and the phases for which quorums of sent messages should be received in order to have a linearization point of $\add$ or $\commit$.
%
A summary is presented in Table~\ref{tb:summary}. The * on the total number of phases means that the first phase is skipped in rounds where the leader is stable. For Multi-Paxos and Raft, if the first phase is skipped, then the linearization point of an $\add$ is determined by a quorum of received messages sent in the next phase (and coincides with the linearization point of a $\commit$). We use ``1/2'' to denote this fact.
In PBFT and HotStuff, due to Byzantine processes, quorums of messages sent in two consecutive phases need to be received in order to ensure that the processes are going to vote on the same valid proposal. The 3rd phase in HotStuff is used to ensure progress and can be omitted when reasoning only about safety. 
\begin{table}
\centering
\caption{Summary of linearization point definitions. For each protocol, we give the total number of phases in a round and the number of the phase for which a quorum of sent messages should be received in order to have a linearization point of $\add$ or $\commit$.} 
	\begin{tabular}{|c|c|c|c|c|}
		\hline
		Class & Protocol & \#Phases & $\add$ Quorum Pha. & $\commit$ Quorum Pha. \\
		\hline
		Single-Decree Cons.& Paxos & 2 & 1 & 2 \\
		\hline
		\multirow{2}{*}{Multi-Decree Cons.} & Multi-Paxos & 2* & 1/2 & 2 \\
		& PBFT & 3* & 2 & 3 \\
		\hline
		\multirow{2}{*}{State Machine Repl.} & Raft & 2* & 1/2 & 2 \\
		& HotStuff & 4 & 2 & 4 \\
		\hline 
	\end{tabular}
	\vspace{1mm}

	\label{tb:summary}
\end{table}


\section{Conclusion and Related Work}\label{sec:related}

We have proposed a new methodology for proving safety of consensus or state-machine replication protocols, which relies on a novel abstraction of their dynamics. This abstraction is defined as a sequential QTree object whose state represents a global view of a protocol execution. The operations of QTree construct a tree structure and model agreement on values or a sequence of state-machine commands as agreement on a fixed branch in the tree. Our methodology applies uniformly to a range of protocols like (multi-)Paxos, HotStuff, Raft, and PBFT. We believe that this abstraction helps in improving the understanding of such protocols and writing correct implementations or optimizations thereof. 

As a limitation, it is not clear whether QTree applies to protocols such as Texel~\cite{DBLP:journals/corr/abs-1908-10716} which do not admit a decomposition in rounds. As future work, we might explore the use of QTree in reasoning about liveness. This would require some fairness condition on infinite sequences of add/commit invocations, and a suitable notion of refinement which ensures that infinite sequences of protocol steps cannot be mapped to infinite sequences of stuttering QTree steps.

The problem of proving the correctness of such protocols has been studied in previous work. 
We give an overview of the existing approaches that starts with safety proof methods based on refinement, which are closer to our approach.

\vspace{1mm}
\noindent
\textbf{Refinement based safety proofs.} Verdi \cite{DBLP:conf/pldi/WilcoxWPTWEA15} is a framework for implementing and verifying distributed systems that contains formalizations of various network semantics and failure models. Verdi provides \emph{system transformers} useful for refining high-level specifications to concrete implementations. As a case study, it includes a  fully-mechanized correctness proof of Raft~\cite{DBLP:conf/cpp/WoosWATEA16}. This proof consists of 45000 lines of proof code (manual annotations) in the Coq language for a 5000 lines RAFT implementation, showing the difficulty of reasoning on consensus protocols and the manual effort required. Iron Fleet~\cite{DBLP:conf/sosp/HawblitzelHKLPR15} uses TLA~\cite{DBLP:books/aw/Lamport2002} style transition-system specifications and refine them to low-level implementations described in the Dafny programming language \cite{DBLP:conf/lpar/Leino10}. 
Boichat et al.~\cite{DBLP:journals/sigact/BoichatDFG03} defines a class of specifications for consensus protocols, which are more abstract than QTree and can make correctness proofs harder. Proving Paxos in their case is reduced to a linearizability proof towards an abstract specification, which is quite complex because the linearization points are \emph{not fixed}, they depend on the future of an execution. As a possibly superficial quantitative measure, their Paxos proof reduces to 7 lemmas that are formalized by Garcia-Perez et al.~\cite{DBLP:conf/esop/Garcia-PerezGMS18,DBLP:journals/corr/abs-1802-05969} in 12 pages (see Appendix B and C in~~\cite{DBLP:journals/corr/abs-1802-05969}), much more than our QTree proof. Our refinement proof is also similar to a linearizability proof, but the linearization points in our case are \emph{fixed} (do not depend on the future of an execution) which brings more simplicity. In principle, the specifications in~\cite{DBLP:journals/sigact/BoichatDFG03} could apply to more protocols, but we are not aware of such a case.
%
%
The inductive sequentialization proof rule~\cite{DBLP:conf/pldi/KraglEHMQ20} 
is used for a fully mechanized correctness proof of a realistic Paxos implementation. This implementation is proved to be a refinement of a \emph{sequential} program which is quite close to the original implementation, much less abstract than QTree, and relies on commutativity arguments implied by the communication-closed round structure~\cite{DBLP:journals/scp/ElradF82}. A similar idea is explored in~\cite{DBLP:journals/pacmpl/GleissenthallKB19}, but in a more restricted context.

%


\vspace{1mm}
\noindent
\textbf{Inductive invariant based safety proofs.} Ivy \cite{DBLP:conf/pldi/PadonMPSS16} is an SMT-based safety verification tool that can be used for verifying inductive invariants about global states of a distributed protocol. In order to stay in a decidable fragment of first-order logic, both the modeling and the specification language of IVY are restricted. A simple model of Paxos obeying these restrictions is proven correct in~\cite{DBLP:journals/pacmpl/PadonLSS17}.

\vspace{1mm}
\noindent
\textbf{Beyond safety.} The TLA+ infrastructure~\cite{DBLP:books/aw/Lamport2002} of Lamport has been used to verify both safety and liveness (termination) of several variations of Paxos, e.g., Fast Paxos~\cite{DBLP:journals/dc/Lamport06} or Multi-Paxos~\cite{DBLP:conf/fm/ChandLS16}. Bravo et al.~\cite{DBLP:conf/wdag/BravoCG20} introduce a generic synchronization mechanism for round changes, called the view synchronizer, which guarantees liveness for various Byzantine consensus protocols including our cases studies HotStuff and PBFT. This work includes full correctness proofs for single-decree versions of HotStuff and PBFT and a two-phase version of HotStuff. PSync~\cite{DBLP:conf/popl/DragoiHZ16} provides a partially synchronous semantics for distributed protocols assuming communication-closed rounds in the Heard-Of model~\cite{DBLP:journals/ijsi/Charron-BostM09}. PSync is used to prove both safety and liveness of a Paxos-like consensus protocol called \emph{lastVoting}.

\vspace{1mm}
\noindent
\textbf{Relating different consensus protocols.} Lamport defines a series of refinements of Paxos that leads to a Byzantine fault tolerant version, which is refined by PBFT~\cite{DBLP:conf/wdag/Lamport11a}. Our proof that Paxos refines QTree can be easily extended to this Byzantine fault tolerant version in the same manner as we did for PBFT.
Wang et al.~\cite{DBLP:conf/podc/WangZ0C019} shows that a variation of RAFT is a refinement of Paxos, which enables porting some Paxos optimizations to RAFT. 
Renesse et al.~\cite{DBLP:journals/tdsc/RenesseSS15} compare Paxos, 
Viewstamped Replication \cite{DBLP:conf/podc/OkiL88} and ZAB~\cite{DBLP:conf/dsn/JunqueiraRS11}. They define a rooted tree of specifications represented in TLA style whose leaves are concrete protocols. Each node in this tree is refined by its children. Common ancestors of concrete protocols show similarities whereas conflicting specifications show the differences. Similarly, \cite{DBLP:conf/icdcn/SongRSD08} shows that Paxos, Chandra-Toueg \cite{DBLP:journals/jacm/ChandraT96} and Ben-Or \cite{DBLP:conf/podc/Ben-Or83} consensus algorithms share common building blocks. Aublin et al.~\cite{DBLP:journals/tocs/AublinGKQV15} propose an abstract data type for specifying existing and possible future consensus protocols. Unlike our QTree, core components of this data type are not implemented and intentionally left abstract so that it can adapt to different network and process failure models. 

\bibliographystyle{splncs04}
\bibliography{dblp,misc}

\newpage
\appendix

\section{Paxos}
\label{app:paxos}

\begin{figure}[t]
\centering
	\includegraphics[width=1\linewidth]{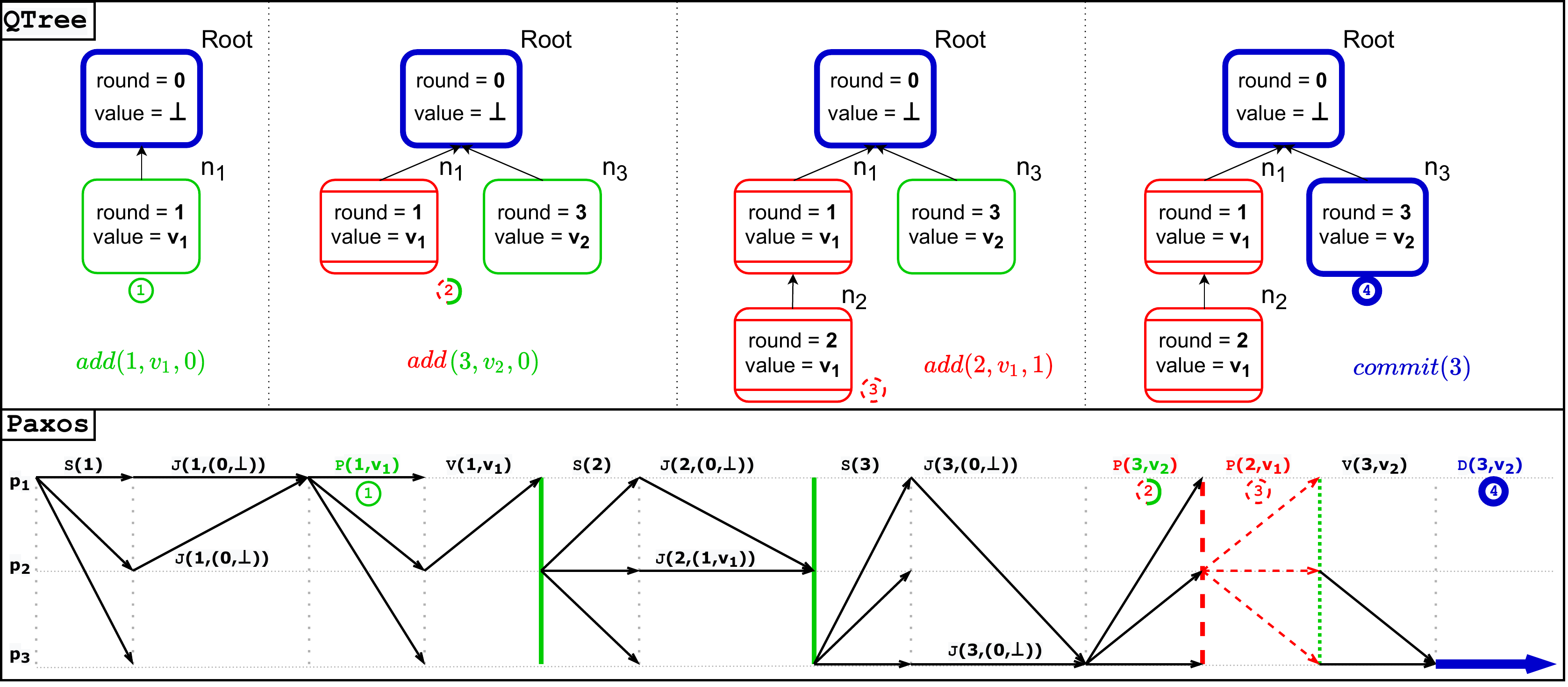}
	\caption{\textbf{Top:} QTree - Explaining the behavior of $\add$ and $\commit$ methods. Colors should be interpreted as in Figure~\ref{fig:qtree}.\\ 
	\textbf{Bottom:} Paxos - A single-decree Paxos execution simulated by the QTree execution above. We abbreviate types of messages with their first letter, e.g., {\fontfamily{qcr}\selectfont START} with {\fontfamily{qcr}\selectfont S}, and payloads are given in parenthesis. Concerning the vertical lines, green solid ones represent the beginning of a new round, red dashed ones show a round that is lagging behind (delayed) and green dotted ones represent returning to a newest round from a delayed round.}
\label{fig:allP}
\vspace{-4mm}
\end{figure}

We illustrate the mapping between linearization points in Paxos and QTree using Paxos execution in Figure~\ref{fig:allP} (Paxos):
\begin{itemize}
\item The leader $p_1$ of round 1 starts the round by broadcasting a {\fontfamily{qcr}\selectfont START}$(1)$ message; $p_1$ and $p_2$ reply with {\fontfamily{qcr}\selectfont JOIN} messages containing empty payloads as they have not sent a {\fontfamily{qcr}\selectfont VOTE} yet. Since $p_1$ receives a quorum of {\fontfamily{qcr}\selectfont JOIN} messages and there is no highest voted round yet, $p_1$ selects a random value ($v_1$), and broadcasts {\fontfamily{qcr}\selectfont PROPOSE}$(1)$. In this step, the linearization point of $\add(1, v_1, 0)$ occurs and it is simulated in QTree with an invocation that adds node $n_1$ with status \texttt{ADDED}. Only $p_2$ replies with a {\fontfamily{qcr}\selectfont VOTE} message, and all the other messages sent in this round are lost. 
	\item The leader $p_2$ of round 2 initiates the round. Only $p_1$ and $p_2$ reply with {\fontfamily{qcr}\selectfont JOIN} messages. Since $p_2$ has voted in round 1, it sends 
the round and the value of the vote. The leader $p_2$ is slow and will resume later.
	\item The leader $p_3$ initiates round 3, and only $p_1$ and $p_3$ reply with {\fontfamily{qcr}\selectfont JOIN} messages that contain empty payloads (they have not voted yet). Hence, $p_3$ selects a random value $v_2$ and broadcasts {\fontfamily{qcr}\selectfont PROPOSE}$(3)$. Here, the linearization point of $\add(3, v_2, 0)$ occurs and it is simulated with a QTree invocation that results in changing the status of $n_1$ to \texttt{GHOST} and adding a new node $n_3$ with status \texttt{ADDED}.
	\item Process $p_2$ resumes and since it received a quorum of {\fontfamily{qcr}\selectfont JOIN} messages in round 2, $p_2$ broadcasts {\fontfamily{qcr}\selectfont PROPOSE}$(2)$ by selecting $v_1$ as the value of the highest voted round from the {\fontfamily{qcr}\selectfont JOIN} messages. At this point, the linearization point of $\add(2, v_1, 1)$ occurs and it is simulated by the QTree invocation that adds node $n_2$ with status \texttt{GHOST}. 
	\item Processes $p_1$ and $p_3$ continue by voting for the proposal in round 3. As a quorum of {\fontfamily{qcr}\selectfont VOTE} messages is received by $p_3$, it decides on $v_2$. Now, the linearization point of $\commit(3)$ occurs which is simulated by changing the status of $n_3$ to \texttt{COMMITTED} in QTree.
\end{itemize}

\section{HotStuff}
\subsection{Complete Description}
\label{ssec:HotStuffdescript}

In the first phase, the leader broadcasts a {\fontfamily{qcr}\selectfont PROPOSE} message to all the processes with a node by executing the {\bf \fontfamily{qcr}\selectfont PROPOSE} action below. This nodes value is sent by a client whose modeling we omit for simplicity. Parent of this node is obtained from processes' {\fontfamily{qcr}\selectfont ROUND-CHANGE} messages. Processes acknowledge with a {\fontfamily{qcr}\selectfont JOIN} message if some conditions are met, executing a {\bf \fontfamily{qcr}\selectfont JOIN} action below: 
\begin{description}
\item[$\bullet$ {\fontfamily{qcr}\selectfont PROPOSE} Action:] \ \\When a proposer $p$ who is the leader of the new round $r$, receives a quorum ($2f + 1$) of {\fontfamily{qcr}\selectfont ROUND-CHANGE}($r$) with $preNode$, it selects the node with the highest round from this set of $preNode$'s. Then it extends the selected node by a newly created node which is initialized with the current round and some value. If there is no such highest round, then the proposer extends the Genesis Block. Finally, proposer $p$ broadcasts {\fontfamily{qcr}\selectfont PROPOSE}($r$) to all processes alongside with the new node.
  
\item[$\bullet$ {\fontfamily{qcr}\selectfont JOIN} Action:] \ \\ When a process $p'$ receives a {\fontfamily{qcr}\selectfont PROPOSE}($r$) with node $n$, if $n$ extends $votedNode$ or the round of the $votedNode$ is less than the round of $n.parent$, then $p'$ sends a {\fontfamily{qcr}\selectfont JOIN}($r$) to the leader of the current round.
\end{description}

If the leader receives acknowledgement messages from a quorum of processes, the second phase starts. The leader broadcasts a {\fontfamily{qcr}\selectfont PRECOMMIT} message with the same node by executing {\bf \fontfamily{qcr}\selectfont PRECOMMIT} action and accepters acknowledge with the {\fontfamily{qcr}\selectfont PRECOMMIT\_VOTE} message, executing {\bf \fontfamily{qcr}\selectfont PRECOMMIT\_VOTE} action.
\begin{description}
\item[$\bullet$ {\fontfamily{qcr}\selectfont PRECOMMIT} Action:] \ \\ When the proposer $p$ receives a quorum of {\fontfamily{qcr}\selectfont JOIN}($r$) with its current round and the same node, $p$ combines (generates certificate) and sends them by broadcasting a {\fontfamily{qcr}\selectfont PRECOMMIT}($r$) with the same node $n$ to all processes.
\item[$\bullet$ {\fontfamily{qcr}\selectfont PRECOMMIT\_VOTE} Action:] \ \\ When a process $p'$ receives a {\fontfamily{qcr}\selectfont PRECOMMIT}($r$) from the leader of its current round, $p'$ updates $preNode$ with the node $n$ that it received and sends a {\fontfamily{qcr}\selectfont PRECOMMIT\_VOTE}($r$) with the same node to the leader of the current round.
\end{description}

Like the previous phase, if the leader receives acknowledgement messages from a quorum of processes, the third phase starts. The leader broadcasts a {\fontfamily{qcr}\selectfont COMMIT} message with the same node by executing {\bf \fontfamily{qcr}\selectfont COMMIT} action and accepters acknowledge with the {\fontfamily{qcr}\selectfont COMMIT\_VOTE} message, executing {\bf \fontfamily{qcr}\selectfont COMMIT\_VOTE} action. Then if the leader receives {\fontfamily{qcr}\selectfont COMMIT\_VOTE} messages from a quorum of processes, the proposed value becomes decided (and sent to the client) by executing a {\bf \fontfamily{qcr}\selectfont DECIDE} action.
\begin{description}
\item[$\bullet$ {\fontfamily{qcr}\selectfont COMMIT} Action:] \ \\ When the proposer $p$ receives a quorum of {\fontfamily{qcr}\selectfont PRECOMMIT\_VOTE}($r$) with its current round and the same node, $p$ combines and sends them by broadcasting a {\fontfamily{qcr}\selectfont COMMIT}($r$) with the same node $n$ to all processes.
\item[$\bullet$ {\fontfamily{qcr}\selectfont COMMIT\_VOTE} Action:] \ \\ When a process $p'$ receives a {\fontfamily{qcr}\selectfont COMMIT}($r$) from the leader of its current round, $p'$ updates $votedNode$ with the node $n$ that it received and sends a {\fontfamily{qcr}\selectfont COMMIT\_VOTE}($r$) with the same node to the leader of the current round.

\item[$\bullet$ {\fontfamily{qcr}\selectfont DECIDE} Action:] \ \\ When the proposer $p$ receives a quorum of {\fontfamily{qcr}\selectfont COMMIT\_VOTE}($r$) with its current round and the same node, $p$ combines and sends them by broadcasting a {\fontfamily{qcr}\selectfont DECIDE}($r$) with the same node $n$ to all processes. When a process $p'$ receives a {\fontfamily{qcr}\selectfont DECIDE}($r$) from the leader of its current round, $p'$ updates $decidedNode$ as $n$ and execute commands through the branch where the leaf node is $n$.
\end{description}

If timeout is reached for a process, {\bf \fontfamily{qcr}\selectfont ROUND-CHANGE} action will be executed.
\begin{description}
\item[$\bullet$ {\fontfamily{qcr}\selectfont ROUND-CHANGE} Action:] \ \\ When the timeout is reached, a process $p'$ sends a {\fontfamily{qcr}\selectfont ROUND-CAHNGE}($r$) with $preNode$ to the leader of the next round. Additionally, $p'$ increments its round number.

\end{description}

\subsection{HotStuff Refines QTree}
\label{ssec:HotStuffproofProps}
\begin{proof}
We show that the sequence of successful $\add$ and $\commit$ invocations defined by linearization points along a HotStuff execution satisfies the properties in Theorem~\ref{th:main} and therefore, it represents a correct QTree execution:
\begin{itemize}
\item \textbf{Property~\ref{eq:prop1}:} To generate a valid (certified under threshold signatures) {\fontfamily{qcr}\selectfont PRECOMMIT}($r$) (resp., {\fontfamily{qcr}\selectfont DECIDE}($r$)), a leader must collect a quorum of {\fontfamily{qcr}\selectfont JOIN}($r$) (resp., {\fontfamily{qcr}\selectfont COMMIT\_VOTE}($r$)) messages with the same content i.e., the same node with the same client request, connected to the same parent. As all the correct replicas will send at most one message per phase in a single round $r$, there can't be two quorums of {\fontfamily{qcr}\selectfont JOIN}($r$) (resp., {\fontfamily{qcr}\selectfont COMMIT\_VOTE}($r$)) resulting two {\fontfamily{qcr}\selectfont PRECOMMIT}($r$) with different contents. Since the linearization point of $\add(r,\_,)$ (resp., $\commit(r)$) occurs when the $first$ valid {\fontfamily{qcr}\selectfont PRECOMMIT}(r) (resp., {\fontfamily{qcr}\selectfont DECIDE}($r$)) message is broadcasted, property holds by the definition. 
\item \textbf{Property~\ref{eq:prop0}:} This holds trivially as there won't be a quorum of {\fontfamily{qcr}\selectfont COMMIT\_VO}-{\fontfamily{qcr}\selectfont TE}($r$) messages without a quorum of {\fontfamily{qcr}\selectfont PRECOMMIT\_VOTE}($r$)  messages which do not exist as there is no valid {\fontfamily{qcr}\selectfont PRECOMMIT}(r).
\item \textbf{Property~\ref{eq:prop2}:} By the definition of {\bf \fontfamily{qcr}\selectfont JOIN} action, a correct process will accept a {\fontfamily{qcr}\selectfont PROPOSE(r)} message if the node that is sent alongside with this message is extending some $preNode$ which can be certified only if a quorum of {\fontfamily{qcr}\selectfont JOIN}($r'$) messages (that forms a {\fontfamily{qcr}\selectfont PRECOMMIT}($r'$)) are sent to the leader for some round $r' > 0$. Since a quorum of {\fontfamily{qcr}\selectfont JOIN}($r'$) messages and {\fontfamily{qcr}\selectfont PROPOSE(r)} message are formed before and after a quorum of {\fontfamily{qcr}\selectfont ROUND-CHANGE}($r$) respectively, $r > r'$. Note that processes can only vote for their current round and the round number monotonically increases. Therefore, to reach a quorum of {\fontfamily{qcr}\selectfont JOIN}($r$) (which is imperative to generate {\fontfamily{qcr}\selectfont PRECOMMIT}($r$)), {\fontfamily{qcr}\selectfont PRECOMMIT}($r'$) must exists.
	\begin{itemize}
        \item \textbf{Property~\ref{eq:prop2a}:} This property doesn't hold (and not needed) for HotStuff.
        \end{itemize} 
\item \textbf{Property~\ref{eq:prop3}:} Assume by contradiction that $\commit(r)$ occured along with the other two linearization points of $\add$. Linearization point of $\commit(r)$ exists because of a quorum of {\fontfamily{qcr}\selectfont COMMIT\_VOTE}($r$) messages sent by a set of processes $P_1$, and $\add(r',\_,r'')$ exists because of a quorum of {\fontfamily{qcr}\selectfont JOIN}($r'$) messages sent by a set of processes $P_2$. All the correct processes in $P_1$ must updated their $\mathit{votedNode}$ with a node whose round is $r$ when they sent {\fontfamily{qcr}\selectfont COMMIT\_VOTE}($r$) message. But also, all the correct processes in $P_2$ must have a $\mathit{votedNode}$ whose round number is less than or equal to $r''$ by the predicates in the definition of the {\bf {\fontfamily{qcr}\selectfont JOIN(r)}} action. Note that none of the correct processes in $P_2$ can send {\fontfamily{qcr}\selectfont COMMIT\_VOTE}($r$) message anymore since their current round number is at least $r'$ which is greater than $r$. Since $P_1$ and $P_2$ must have an intersecting correct process, it contradicts the hypothesis as $r''<r$.
\end{itemize}
\end{proof}
\section{Raft}
\label{appendix:raft}
\subsection{Complete Description}
Raft is a partially synchronous, multi-decree consensus protocol that is resilient to only crash-restart failures. Each process of Raft keeps a durable $log$ for storing the sequence of commands. 

Raft has the notion of \emph{terms} that does not exactly match with our round notion. For each term, there is at most one leader that does not change throughout the term.  When the current term's leader is suspected to fail, processes start a new leader election for a new bigger term number. The leader might propose values for different commands (log indices) inside a term as long as it stays alive. In order to differentiate values proposed and decided for distinct commands within a term, we keep term - index pairs ($r = (t, idx)$) as rounds. We assume the usual lexicographical total ordering on rounds. 

Logs keep term - value pairs at each index. Here, the first field represents the term at which this element is created and inserted to the log and the second field represents the command offered for this index. For each process, a prefix of the $log$ is called decided. If it is decided, then this prefix is supposed to be the same for a majority of process logs. Remaining parts of the logs (suffixes) might be different among processes. Length of the uncommon suffix might be more than one and different between processes since a leader might append multiple items to the log at once and these new entries might arrive to a subset of processes. For each process, we keep two special indices: $didx_p$ marks the end of the decided prefix whereas $lidx_p$ shows the last entry's index for the log of process $p$. 

Raft rounds consist of a single main phase ignoring the leader election phase that does not happen at every round. Leader election phases are only executed during term changes. The leader election phase can be considered as the first phase of the first round of the new term.

When a process $p$ suspects from the inactivity of the current leader, it broadcasts {\fontfamily{qcr}\selectfont VOTEREQ} message to initiate the leader election phase and it becomes the candidate leader for the new term. When a process receives this message, it responds to $p$ with a {\fontfamily{qcr}\selectfont VOTERESP} message if some conditions are met. If $p$ can get {\fontfamily{qcr}\selectfont VOTERESP} messages from a majority, it becomes the leader of the new term.

\begin{description}
	\item [$\bullet$ {\bf \fontfamily{qcr}\selectfont VOTEREQ} Action:] \ \\ This action is executed when process $p$ times out while waiting for a message from the leader of the term $t-1$. Process $p$ broadcasts {\fontfamily{qcr}\selectfont VOTEREQ($t, lidx_p$)} message and updates its term $term_p$ to $t$.
	\item [$\bullet$ {\bf \fontfamily{qcr}\selectfont VOTERESP} Action:] This action can only be executed by process $p'$ after some {\fontfamily{qcr}\selectfont VOTEREQ($t, lidx_p$)} action from the candidate leader $p$. With this action, $p'$ sends the message {\fontfamily{qcr}\selectfont VOTERESP($t, lidx_{p'}$)} to $p$ and updates its term $term_{p'}$ to $t$ if (1) it has not sent any {\fontfamily{qcr}\selectfont VOTERESP} message for the term $t$ or higher before ($t \geq term_{p'}$), and (2) $(log_{p'}[lidx_{p'}].term, lidx_{p'}) \leq (log_p[lidx_p].term, lidx_p)$. Second condition means that the last item in $p$'s log has been proposed in a bigger or the same round than the last item in the log of $p'$. 
\end{description}

If $p$ can collect {\fontfamily{qcr}\selectfont VOTERESP} messages from a majority, $p$ becomes the leader of term $t$ and start proposing values. Since a process can send at most one {\fontfamily{qcr}\selectfont VOTERESP} message for any term, there is at most one leader for each term.

After the leader is elected, it starts sending requests to processes to append new values to their logs by iterating the main phase of rounds. When new commands come from clients,  the leader $p$ first appends them to its own log with the current term, increments $lidx_p$ and then broadcasts {\fontfamily{qcr}\selectfont LOGREQ} messages that include the new entries. When a process $p'$ receives this message, it checks some conditions. If conditions are satisfied, it updates its $didx_{p'}$, $lidx_{p'}$ and $log_{p'}$ and then responds with a {\fontfamily{qcr}\selectfont LOGRESP} message to the leader. If the leader receives a {\fontfamily{qcr}\selectfont LOGRESP} message from a majority, it confirms that the new commands became permanent in a majority of processes and updates its $didx$ value. 

\begin{description}
	\item [$\bullet$ {\bf \fontfamily{qcr}\selectfont LOGREQ} Action:] \ \\ This action is executed by the leader process $p$ of $term_p$. If this action is not the first {\bf \fontfamily{qcr}\selectfont LOGREQ} action of this term, the leader first checks if there is a set of {\fontfamily{qcr}\selectfont LOGRESP} messages from a majority for the previous {\bf \fontfamily{qcr}\selectfont LOGREQ} action. If this is the case, it updates $didx_p$ value to $lidx_p$ and decides on the entries appended in the previous turn. Then, it appends new entries to $log_p$ and updates $lidx_p$ so that it now points to the end of $log_i$. As the last thing, it broadcasts {\fontfamily{qcr}\selectfont LOGREQ($t, lidx_p$)} message with its $log_p$ and $didx_p$. 
	\item [$\bullet$ {\bf \fontfamily{qcr}\selectfont LOGRESP} Action:] \ \\ This action can be only executed by process $p'$ after receiving a {\fontfamily{qcr}\selectfont LOGREQ} message. First $p'$ checks whether $term_p \geq term_{p'}$ and $lidx_p \geq lidx_{p'}$. If this is the case, it updates $term_{p'}$, $lidx_{p'}$ and $didx_{p'}$ to $term_p$, $lidx_p$ and $didx_p$, respectively. Moreover, for each index $i$ in until and including $lidx_i$, it replaces $log_{p'}[i]$ with $log_p[i]$. Then it sends  {\fontfamily{qcr}\selectfont LOGRESP($t, lidx_{p'}$)} response back to the leader process $p$.
\end{description}

Even if the leader $p$ does not receive a new value from the clients for a long time, it still broadcasts a {\fontfamily{qcr}\selectfont LOGREQ} message with the same $log_p$ and $lidx_p$ value as the previous {\fontfamily{qcr}\selectfont LOGREQ} message to signal to other processes that it is still alive. These {\fontfamily{qcr}\selectfont LOGREQ} messages are called \emph{heartbeat} messages. They can only differ on $didx_p$ values since a majority quorum might send {\fontfamily{qcr}\selectfont LOGRESP} messages in between two heartbeat messages that can changed the $didx_p$ values. We also include heartbeat messages in our formulation.

\subsection{Raft Refines QTree}

Our main correctness theorem for Raft is as follows:

\begin{theorem}\label{th:raft}
	Raft refines QTree.
\end{theorem}

As the first step towards the proof of Theorem~\ref{th:raft}, we will determine the Raft actions that will correspond to linearization points of successful $\add$ and $\commit$ invocations.

Inside a term $t$, if the leader $p$ receives a {\fontfamily{qcr}\selectfont LOGRESP} message from a majority, this quorum becomes a witness for the decision of entries in the log and proposal of the new entries that will be coming from the clients. In some sense, they correspond to {\fontfamily{qcr}\selectfont VOTE} and {\fontfamily{qcr}\selectfont JOIN} quorums of Paxos, respectively. When the leader broadcasts a {\fontfamily{qcr}\selectfont LOGREQ} message first time in a new term, the witness quorum for the proposal of new entries is formed by {\fontfamily{qcr}\selectfont VOTERESP} messages received from a majority. Therefore, both $\add$ and $\commit$ linearization points correspond to the {\bf \fontfamily{qcr}\selectfont LOGREQ} actions. 

Assume that $old\_didx_p$ to represent the $didx_p$ value before executing the {\bf \fontfamily{qcr}\selectfont LOGREQ} action. Then,
\begin{itemize}
\item the linearization point of $add((log_p[i].term, i),\allowbreak log_p[i].value, (log_p[i-1].term,$ $i-1))\Rightarrow OK$ occurs in \textbf {\fontfamily{qcr}\selectfont LOGREQ} action for all indices $i$ such that $lidx_p \geq i > didx_p$ and $log_p[i].term = term_p$. For the case $i = 0$, we replace the round $(log_p[i-1].term, i-1)$ with $\bot$.
\item the linearization point of $commit( (log_p[i].term, i))\Rightarrow OK$ occurs for (again during  \textbf{\fontfamily{qcr}\selectfont LOGREQ} action) for all indices $i$ such that $didx_p \geq i > old\_didx_p$ and $log_p[i].term = term_p$. Note that $old\_didx_p$ is not defined if \textbf{ \fontfamily{qcr}\selectfont LOGREQ} is the first such action of $term_p$. Indeed, this action is not a $\commit$ linearization point in this case.
\end{itemize}

Next, in order to prove Theorem~\ref{th:raft}, we show that sequence of successful $\add$ and $\commit$ invocations, defined by linearization points along a Raft execution satisfies the properties in Theorem~\ref{th:main}.

Before explaining why Raft executions satisfy our properties, we introduce two additional properties of Raft that will be used during the proofs.
\begin{enumerate}
	\item \label{RaftProp1} Consider a $log$ of a process in a reachable Raft state. For any two indices $i \leq i' \leq lidx$, we have $log[i].term \leq log[i'].term$. 
	\item \label{RaftProp2} Assume that $commit(r)$ action is generated for some $r = (t, i)$. Now, consider the log of a leader $p$ for some term $t' \geq t$ during this term $t'$. We have $log_p[i].term = t$.
\end{enumerate}

Raft's leader election mechanism ensures that there is a unique leader that can execute {\bf \fontfamily{qcr}\selectfont LOGREQ} action and append entries to logs at any time and the term of the current leader is bigger than all previously active terms. These ensure property~\ref{RaftProp1}.

During the leader election process, processes with the longest logs are the ones that received the latest updates  from the previous leader. Since {\fontfamily{qcr}\selectfont LOGRESP} and {\fontfamily{qcr}\selectfont VOTERESP} quorums intersect, if there is a decided entry in one of the previous terms, the new leader has the same entry in its log as well. This ensures property~\ref{RaftProp2}.

The proof of the properties for the linearization points along a Raft execution is as the following:
\begin{itemize}
\item \textbf{Property~\ref{eq:prop1}:} Consider any round $r = (t, i)$. Leader election phase of Raft ensures that a unique leader can execute {\bf \fontfamily{qcr}\selectfont LOGREQ} actions inside the term $t$. Moreover, log of the leader only grows and items in previously entered indices never change through a term. Therefore, there is a unique $add(r,\_, \_)$ linearization point for each round $r$. There is a unique $commit(r)$ linearization point due to previously mentioned properties of the leader and $didx$ of the leader is non-decreasing through a term.
\item \textbf{Property~\ref{eq:prop0}:} Consider a $commit(r)$ linearization point for some $r = (t, i)$. Since $commit(r)$ linearization action exists, {\bf \fontfamily{qcr}\selectfont LOGREQ} action that leads to this point is not the first {\bf \fontfamily{qcr}\selectfont LOGREQ} action of this term. Moreover, the last non-heartbeat {\bf \fontfamily{qcr}\selectfont LOGREQ} action before this one is the linearization point for $add(r,\_,\_)$.
\item \textbf{Property~\ref{eq:prop2}:} Consider an $add(r, v, r')$ linearization point where $r = (log[i].$\\$term, i)$ and $r' = (log[i-1].term, i-1)$ for some $i > 0$. Property~\ref{RaftProp1} ensures that $log[i].term \geq log[i-1].term$. Therefore $r > r'$ according to the lexicographical ordering we have on rounds. Moreover, since there is an item in $log[i-1]$, there must be a {\bf \fontfamily{qcr}\selectfont LOGREQ} action that caused this value to be inserted into a log first time. This action must have led to $add(r', v', \_)$ linearization point.
	\begin{itemize}
        \item \textbf{Property~\ref{eq:prop2a}:} This property does not hold for Raft.
        \end{itemize} 
\item \textbf{Property~\ref{eq:prop3}:} Towards a contradiction, assume that there are $add(r,\_, \_)$, $add(r',\_, r'')$ and $commit(r, \_)$ linearization points where $r = (t, i)$, $r' = (t', i')$ and $r'' = (t'', i'')$. Total ordering on rounds ensures that $t' \geq t \geq t''$. In terms of the second fields, the second $add$ linearization point ensures that $i' = i'' + 1$.

Consider the log of the leader of the term $t'$ that generates the second $add$ at the state it generated this linearization point. Since there is a $commit(r,\_)$ linearization point, property~\ref{RaftProp2} ensures that $log[i].term = t$. Moreover, we have $log[i'].term = t'$ and $log[i''] = log[i'-1].term = t''$. Next, we consider different cases on $i$.

First of all $i = i'$ cannot be true. If this was the case, since $t = log[i].term = log[i'].term = t'$,  we would have $r = r'$. Therefore, we consider $i > i'$ as the first case. For this case, we have $t' > t \geq   t''$. This case violates property~\ref{RaftProp1} since $i > i'$ but $term[i] = t < t' = term[i']$.

The last case we consider is $i < i'$. For this case, we have $t' \geq t > t''$. But, again property~\ref{RaftProp1} is violated since $i \leq i''$ but $term[i] = t > t'' = term[i'']$.
\end{itemize}

If we look at the proofs of our properties, the restriction we have on quorums is that {\fontfamily{qcr}\selectfont LOGRESP} and {\fontfamily{qcr}\selectfont VOTERESP} quorums intersect. We do not need two {\fontfamily{qcr}\selectfont LOGRESP} or two {\fontfamily{qcr}\selectfont VOTERESP} quorums to intersect. Therefore, different correct Raft variants can be developed with different quorum sizes. For instance, if the leaders are mostly stable and the network is reliable, one can modify the {\fontfamily{qcr}\selectfont LOGRESP} quorum size to a smaller value so that entries can be appended more efficiently, but {\fontfamily{qcr}\selectfont VOTERESP} quorum size must be increased by the same amount to still enforce safety guarantees. 
\section{PBFT}
\subsection{Complete Description}
\label{ssec:PBFTdescript}

In the first phase, the leader broadcasts a {\fontfamily{qcr}\selectfont PROPOSE} message to all the processes with a value by executing the {\bf \fontfamily{qcr}\selectfont PROPOSE} action below. This value can be a value sent by a client (whose modeling we omit for simplicity) or it can be obtained from processes' {\fontfamily{qcr}\selectfont ROUND-CHANGE} messages that started this round. Processes accept this proposal with broadcasting a {\fontfamily{qcr}\selectfont JOIN} message if some conditions are met, executing a {\bf \fontfamily{qcr}\selectfont JOIN} action below:
\begin{description}
\item[$\bullet$ {\fontfamily{qcr}\selectfont PROPOSE} Action:] \ \\When a proposer $p$ who is the leader of the new round $r$, receives a quorum of {\fontfamily{qcr}\selectfont ROUND-CHANGE}($r$) with certificates, it selects the valid (contains quorum of matching  {\fontfamily{qcr}\selectfont JOIN} alongside) {\fontfamily{qcr}\selectfont VOTE} with the highest round for some available sequence number $sn$ and propose its value by broadcasting a new {\fontfamily{qcr}\selectfont PROPOSE}($r$, $sn$) for the current round $r$. If there is no such highest round, then $p$ selects the proposed value randomly simulating the value coming from the client. The proposer $p$ also sends the set of {\fontfamily{qcr}\selectfont VOTE} messages included in the {\fontfamily{qcr}\selectfont ROUND-CHANGE}($r$) messages it received, to prove that it selected a valid {\fontfamily{qcr}\selectfont VOTE} with the highest round. 
\item[$\bullet$ {\fontfamily{qcr}\selectfont JOIN} Action:] \ \\ When a processor $p'$ receives some number of {\fontfamily{qcr}\selectfont PROPOSE} with its current round, it can only act to one of them if their sequence numbers are the same. During a round, if $p'$ didn't see any {\fontfamily{qcr}\selectfont PROPOSE} with the same sequence number, then it checks whether the proposers selection is correct. After validating that proposer $p$ selected the {\fontfamily{qcr}\selectfont VOTE} with the highest round, $p'$ broadcasts {\fontfamily{qcr}\selectfont JOIN}($r$, $sn$) to all processes using the same content.
\end{description}

If a process receives a quorum of {\fontfamily{qcr}\selectfont JOIN} messages, the second phase starts. The processes broadcast a {\fontfamily{qcr}\selectfont VOTE} message, executing {\bf \fontfamily{qcr}\selectfont VOTE} action. Then if a process receives a quorum of {\fontfamily{qcr}\selectfont VOTE} messages, the proposed value becomes decided for this process (and sent to the client) during the execution of {\bf \fontfamily{qcr}\selectfont DECIDE} action.
\begin{description}
\item[$\bullet$ {\fontfamily{qcr}\selectfont VOTE} Action:] \ \\ When a process $p'$ receives a {\fontfamily{qcr}\selectfont PROPOSE} and a quorum of {\fontfamily{qcr}\selectfont JOIN} with its current round and the same sequence number, it broadcasts a {\fontfamily{qcr}\selectfont VOTE}($r$, $sn$) to all processes using the same values.

\item[$\bullet$ {\fontfamily{qcr}\selectfont DECIDE} Action:] \ \\ When a process $p'$ receives a {\fontfamily{qcr}\selectfont PROPOSE}, a quorum of {\fontfamily{qcr}\selectfont JOIN} and a quorum of {\fontfamily{qcr}\selectfont VOTE} with its current round and the same sequence number, it updates a local variable $decidedVal[sn]$ with the value that it received for the order. This assignment means that the value is decided for the order $sn$ and sent to the client. Since there may be $f$ Byzantine processes, the client accept the decision if it receives $f+1$ of the same decided value for the same sequence number. \end{description}

If timeout is reached for a process, {\bf \fontfamily{qcr}\selectfont ROUND-CHANGE} action will be executed.
\begin{description}
\item[$\bullet$ {\fontfamily{qcr}\selectfont ROUND-CHANGE} Action:] \ \\ When the timeout is reached, a process $p'$ sends a {\fontfamily{qcr}\selectfont ROUND-CAHNGE}($r$) to the proposer of the next round. Additionally, for all the sequence numbers which are not decided by a quorum yet, if $p'$ sent {\fontfamily{qcr}\selectfont VOTE} for some of these sequence numbers before, then $p'$ sends the one with the highest round for each of these sequence numbers. The process $p'$ sends them as a certificate which consists of {\fontfamily{qcr}\selectfont VOTE}, the matching {\fontfamily{qcr}\selectfont PROPOSE} and the quorum of {\fontfamily{qcr}\selectfont JOIN} that it received. Finally, the process $p'$ increments its round number.

\end{description}

\subsection{PBFT Refines A Set of QTree Instances}
\label{ssec:PBFTproofProps}
\begin{proof}
We show that the sequence of successful $\add$ and $\commit$ invocations on a QTree instance $sn$ defined by linearization points along a PBFT execution satisfies the properties in Theorem~\ref{th:main} and therefore, it represents a correct QTree execution:
\begin{itemize}
\item \textbf{Property~\ref{eq:prop1}:} By the definition of {\bf \fontfamily{qcr}\selectfont JOIN} action, a correct process can only send a {\fontfamily{qcr}\selectfont JOIN}($r$, $sn$) message if this process has not sent a {\fontfamily{qcr}\selectfont JOIN} message for the same round and sequence number yet. Hence, there can be at most 1 quorum (2f+1) of {\fontfamily{qcr}\selectfont JOIN} messages with the same sequence number and round since any two quorums must intersect with a correct process. This implies that the linearization point of $sn.\add(r,\_,\_)$ will occur at most once for a round $r$ with the same sequence number $sn$. Therefore, a correct process can only vote for a single propose in a round and only 1 proposal can be decided each round. Thus, at most one linearization point of $sn.\commit(r)$ can occur for a round $r$ with the same sequence number $sn$. 
\item \textbf{Property~\ref{eq:prop0}:} This holds trivially as there won't be a quorum of {\fontfamily{qcr}\selectfont VOTE}($r$, $sn$) messages without a {\fontfamily{qcr}\selectfont VOTE}($r$, $sn$) from an honest process since there is no quorum of {\fontfamily{qcr}\selectfont JOIN}($r$, $sn$) messages.
\item \textbf{Property~\ref{eq:prop2}:} By the definition of {\bf \fontfamily{qcr}\selectfont PROPOSE} action, proposer selects a highest vote round number $r'$ from a quorum of {\fontfamily{qcr}\selectfont ROUND-CHANGE}($r$) messages that it receives for each sequence number, before broadcasting a {\fontfamily{qcr}\selectfont PROPOSE}($r$, $sn$) message. Since all processes also send the quorum of {\fontfamily{qcr}\selectfont JOIN} messages that they receive as a certificate for each {\fontfamily{qcr}\selectfont VOTE} that they send in their {\fontfamily{qcr}\selectfont ROUND-CHANGE} messages, a correct proposer can validate {\fontfamily{qcr}\selectfont VOTE} messages. If the proposer is faulty, it may select a non-existing vote but correct processes do not proceed with the current proposal since it is not valid. If correct processes can progress with the current proposal where there is a highest vote round number $r' > 0$ selected by the proposer, and one of the processes sends a {\fontfamily{qcr}\selectfont VOTE}($r$, $sn$), then there must be a {\fontfamily{qcr}\selectfont VOTE}($r'$, $sn$) message.  Hence, if the linearization point of $sn.\add(r,\_,r')$ occurs in PBFT where $r' \neq 0$, then it is preceded by $sn.\add(r',\_,\_)$. 
	\begin{itemize}
	\item \textbf{Property~\ref{eq:prop2a}:} If the proposer is correct, it must propose the same value
	with the vote that is selected from a quorum of {\fontfamily{qcr}\selectfont ROUND-CHANGE}
	messages by definition. Else if the proposer omit the definition and send a different value, correct
	processes do not accept the current proposal and the quorum will not be formed.Therefore, when the linearization point of both $sn.\add(r,v,r')$ and $sn.\add(r',v',\_)$ occur in a PBFT execution, $\mathit{v = v'}$.
	\end{itemize}
\item \textbf{Property~\ref{eq:prop3}:} Assume by contradiction that $sn.\commit(r)$ occurred along with the other two linearization points of $\add$. The linearization point of $sn.\commit(r)$ occurs because of a quorum of {\fontfamily{qcr}\selectfont VOTE}($r$, $sn$) messages sent by a set of processes $P_1$, and $sn.\add(r',\_,r'')$ occurs because of a quorum of {\fontfamily{qcr}\selectfont JOIN}($r'$, $sn$) messages sent by a set of processes $P_2$. All the correct processes in $P_1$ must send {\fontfamily{qcr}\selectfont VOTE}($r$, $sn$) messages to the leader of the next round therefore they must  send {\fontfamily{qcr}\selectfont VOTE}($r_1$, $sn$) to the leader of round $r'$ where $r \leq r_1 < r'$. But $sn.\add(r',\_,r'')$ shows that during round $r'$, quorum of {\fontfamily{qcr}\selectfont JOIN}($r'$, $sn$) messages could be sent because all the correct processes in $P_2$ accepts that {\fontfamily{qcr}\selectfont VOTE}($r''$, $sn$) is the message with the highest round (in {\fontfamily{qcr}\selectfont ROUND-CHANGE}($r'$) messages). Since $P_1$ and $P_2$ must have a an intersecting correct process and $r'' < r_1$, it is a contradiction.
\end{itemize}
\end{proof}

\begin{figure}[t]
\centering
	\includegraphics[width=1\linewidth]{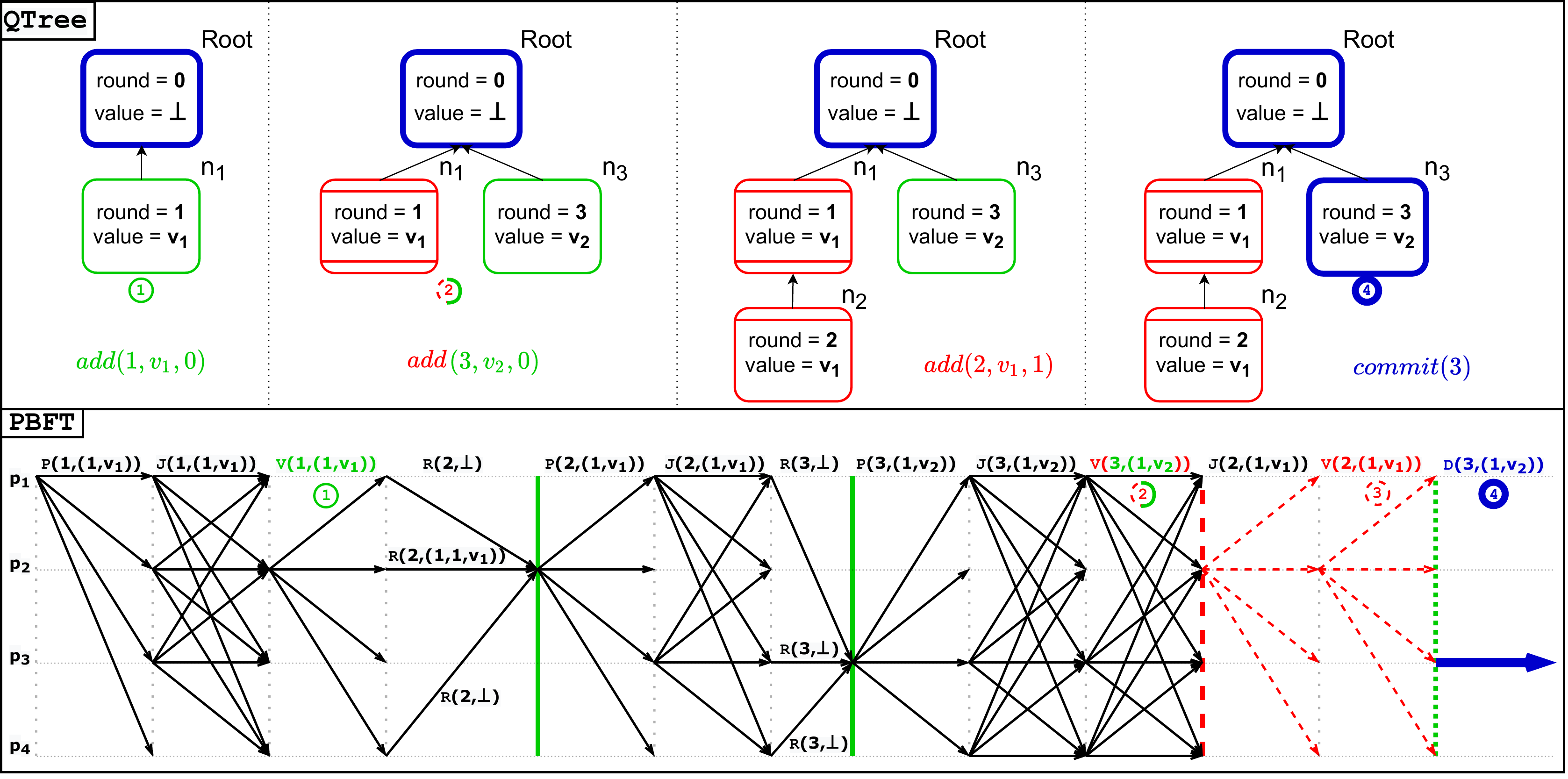}
	\caption{\textbf{Top:} QTree - Explaining the behavior of $\add$ and $\commit$ methods. Colors should be interpreted as in Figure~\ref{fig:qtree}\\ 
	\textbf{Bottom:} PBFT - A PBFT execution (for sequence number 1) simulated by the QTree execution above. We abbreviate types of messages with their first letter, e.g., {\fontfamily{qcr}\selectfont ROUND-CHANGE} with {\fontfamily{qcr}\selectfont R}, and payloads are given in parenthesis. Concerning the vertical lines, colors should be interpreted as in Figure~\ref{fig:all}}. 
	\label{fig:allPBFT}
\end{figure}

\subsection{Explanation of Figure~\ref{fig:allPBFT}(PBFT)}
\label{ssec:PBFTfig}
We illustrate the mapping between protocol steps and QTree steps using the PBFT execution for sequence number 1 in Figure~\ref{fig:allPBFT} (PBFT). The corresponding QTree execution is given just above. Therefore:
\begin{itemize}
	\item The leader $p_1$ of round 1 starts the round by broadcasting a {\fontfamily{qcr}\selectfont PROPOSE}$(1)$ message where $v_1$ is selected randomly; $p_1$, $p_2$ and $p_3$ acknowledge the proposal by broadcasting {\fontfamily{qcr}\selectfont JOIN} messages. Now only $p_2$ broadcasts {\fontfamily{qcr}\selectfont VOTE}$(1,1,v_1)$ message after recieveing a quorum of {\fontfamily{qcr}\selectfont JOIN}. In this step, the linearization point of $1.\add(1, v_1, 0)$ occurs and it is simulated in QTree with an invocation that adds node $n_1$ with status \texttt{ADDED}. 
	\item  When $p_1$, $p_2$ and $p_4$ becomes active, they send {\fontfamily{qcr}\selectfont ROUND-CHANGE} messages to the leader of round 2 which is $p_2$. Here, since $p_2$ is already voted for this sequence number, it sends its vote alongside {\fontfamily{qcr}\selectfont ROUND-CHANGE}, others send empty payloads. Then $p_2$ starts round 2 with sending {\fontfamily{qcr}\selectfont PROPOSE} by selecting $v_1$ as the value of the highest voted round. But due to some connection loss, no more progress can be done in round 2 as the next quorum is not achieved (there is 1 missing message).
	\item After that $p_2$ crashes and so, $p_1$, $p_3$ and $p_4$ send {\fontfamily{qcr}\selectfont ROUND-CHANGE} messages with empty payloads as they haven't voted yet. Hence, $p_3$ (which is the leader of the round 3) selects a random value $v_2$ and broadcasts {\fontfamily{qcr}\selectfont PROPOSE} message for round 3. Later in the same round, $p_1$, $p_3$ and $p_4$ continue by sending {\fontfamily{qcr}\selectfont JOIN} and {\fontfamily{qcr}\selectfont VOTE} messages to the leader of the round. Here, the linearization point of $1.\add(3, v_2, 0)$ occurs and it is simulated with a QTree invocation that results in changing the status of $n_1$ to \texttt{GHOST} and adding a new node $n_3$ with status \texttt{ADDED}
	\item Process $p_2$ becomes active again and since it receives a quorum of {\fontfamily{qcr}\selectfont JOIN} messages in round 2 by sending one final message to itself (which is the current round of $p_2$), $p_2$ broadcasts {\fontfamily{qcr}\selectfont VOTE} message. At this point, the linearization point of $1.\add(2, v_1, 1)$ occurs and it is simulated by the QTree invocation that adds node $n_2$ with status \texttt{GHOST}. 
	\item Finally, as a quorum of {\fontfamily{qcr}\selectfont VOTE} messages is received, $p_3$ decides on $v_2$ and this changes the status of $n_3$ to \texttt{COMMITTED} due to the occurrence of the linearization point of $\commit(3)$.
\end{itemize}
\section{Multi-Paxos (and its variants)} \label{ssec:MultiPaxos}
\subsection{Complete Description}
Multi-Paxos runs a single-decree protocol instance like Paxos concurrently for each sequence number. Since Single-decree Paxos decides on one value, the easy way to agree on sequence of numbers would be to run Paxos multiple times for each sequence number. Multi-Paxos is more efficient version of such an approach. The main optimization in Multi-Paxos is to skip the first phase in Single-decree Paxos and not start a new round when the leader is stable. In other words, after the proposer updates $decidedVal$ with some decided value for sequence number $sn$ in round $r$, it will directly propose some value for $sn+1$ in $r$. When the leader crashes, the next leader starts the round as Single-Decree Paxos but as an optimization, it sends a single {\fontfamily{qcr}\selectfont START} message for each sequence number simultaneously. Then processes send {\fontfamily{qcr}\selectfont JOIN} responses alongside with the highest voted rounds and values for each sequence number. The leader will continue with proposals one by one by selecting highest votes if there exist.

In the first phase, the leader broadcasts a generic {\fontfamily{qcr}\selectfont START} message to all the processes to start the round, executing the  {\bf \fontfamily{qcr}\selectfont START} action below, and processes acknowledge with a {\fontfamily{qcr}\selectfont JOIN} message if some conditions are met, executing the {\bf {\fontfamily{qcr}\selectfont JOIN}} action below:
\begin{description}
\item[$\bullet$ {\fontfamily{qcr}\selectfont START} Action:] The leader $p$ of round $r > 0$ (the proposer) broadcasts a {\fontfamily{qcr}\selectfont START}($r$) message to all processes, waiting their highest votes for each sequence number $sn$.
  
\item[$\bullet$ {\fontfamily{qcr}\selectfont JOIN} Action:] When a process $p'$ receives a {\fontfamily{qcr}\selectfont START}($r$) message, if $p'$ has not sent a {\fontfamily{qcr}\selectfont JOIN} or {\fontfamily{qcr}\selectfont VOTE} message (explained below) for a higher round in the past, it replies by sending a {\fontfamily{qcr}\selectfont JOIN}($r$) message to the proposer. This message includes maximum round numbers ($\mathit{maxVotedRound}$) of all sequence numbers for which $p'$ has sent a {\fontfamily{qcr}\selectfont VOTE} message in the past and the value ($\mathit{maxVotedValue}$) proposed in that round. For each sequence number that it has not voted yet, these fields are $0$ and $\bot$.
\end{description}

If the leader receives {\fontfamily{qcr}\selectfont JOIN} messages from a quorum of processes, the second phase starts. The leader broadcasts {\fontfamily{qcr}\selectfont PROPOSE} message for the next sequence number, executing the {\bf \fontfamily{qcr}\selectfont PROPOSE} action below. 
Processes may acknowledge with a {\fontfamily{qcr}\selectfont VOTE} message if some conditions are met, executing a {\bf \fontfamily{qcr}\selectfont VOTE} action. If the leader receives {\fontfamily{qcr}\selectfont VOTE} messages from a quorum of processes, then the proposed value becomes decided (and sent to the client) by executing a {\bf \fontfamily{qcr}\selectfont DECIDE} action. When the leader decides on some value for sequence number $sn$ without any failure in round $r$, the leader continues with {\fontfamily{qcr}\selectfont PROPOSE} action for $sn + 1$ in $r$. Otherwise a new round is inititated:
\begin{description}
\item[$\bullet$ {\fontfamily{qcr}\selectfont PROPOSE} Action:] When the proposer $p$ receives {\fontfamily{qcr}\selectfont JOIN}($r$) messages from a quorum of ($f + 1$) processes, it selects the one with the highest vote round number for the current sequence number and proposes its value by broadcasting a {\fontfamily{qcr}\selectfont PROPOSE}($r$, $sn$) message (which includes that value). If there is no such highest round (all vote rounds are 0), then the proposer selects the proposed value randomly simulating the value coming from the client (whose modeling we omit for simplicity). 
 
\item[$\bullet$ {\fontfamily{qcr}\selectfont VOTE} Action:] When a process $p'$ receives a {\fontfamily{qcr}\selectfont PROPOSE}($r$, $sn$) message, if $p'$ has not sent a {\fontfamily{qcr}\selectfont JOIN} or {\fontfamily{qcr}\selectfont VOTE} message for a higher round in the past for $sn$, it replies by sending a {\fontfamily{qcr}\selectfont VOTE}($r$, $sn$) message to the proposer with round number $r$ and the same sequence number $sn$. 

\item[$\bullet$ {\fontfamily{qcr}\selectfont DECIDE} Action:] When the proposer $p$ receives {\fontfamily{qcr}\selectfont VOTE}($r$, $sn$)  messages from a quorum of processes, it updates a local variable called $\mathit{decidedVal[sn]}$ to be the value it has proposed in this round $r$ for $sn$. This assignment means that the value is decided and sent to the client after deciding for all sequence number $sn' < sn$.
 
\end{description}

Now we look at the descriptions of protocols which are variants of Multi-Paxos. The protocols that we consider in this section are Cheap Paxos~\cite{DBLP:journals/corr/cs-DC-0408036}, Stoppable Paxos~\cite{malkhi2008stoppable}, Fast Paxos~\cite{DBLP:journals/dc/Lamport06} and Flexible Paxos~\cite{DBLP:journals/corr/HowardMS16}.

Cheap Paxos is a variation of Multi-Paxos where additionally, $f$ of the processes are idle as long as remaining $f+1$ of them are the processes that generated the quorum in the first phase and remain alive. This optimization relies on the fact that, any leader can decide on sequence of values as long as all the processes in a fixed quorum are active. When there is a failure in this fixed quorum, current round ends and the new round starts after the crashed process is replaced with one of the idle process. Since there can be at most $f$ faulty processes, there will be always (at least) one process which will exist in two consecutive quorums, not being idle.

Stoppable Paxos contains special $\mathit{stp}$ command that can be proposed by a leader in some round for a sequence number $sn$ and when this proposal is decided, no more commands are executed for sequence numbers $sn' > sn$. Since this variant enables to stop the current protocol and starts a new one using the final state, a replicated state machine can work as a sequence of stoppable state machines.

In Multi-Paxos, when there is no vote for some sequence number $sn$, the leader receives the value from the client and broadcasts to the processes. In Fast Paxos, when the first phase is skipped as in Multi-Paxos and the leader doesn't receive any vote for the current sequence number, the leader informs clients to send their request directly to all processes rather than to itself. The purpose of this approach is to reduce the end-to-end latency by allowing clients to send their requests directly to the processes but not through the leader (decreasing message delay). Then the processes send $\mathit{fast}$ votes according to the request that they receive and the leader decides on a value if there is a quorum of votes on the same value. In Fast Paxos, a quorum requires $2f+1$ processes where the number of all processes is $3f+1$. When the new round starts, for each sequence number, the leader select highest votes as listed below:
\begin{itemize}
\item If there is not a single vote, the leader selects the value to propose randomly.
\item If there is only a single highest vote, the leader selects the value of that vote.
\item If there are multiple votes, the one which is voted by $f+1$ processes must be selected. If there is no such vote even though there are multiple votes, the leader selects the value of the proposal randomly.
\end{itemize} 

Flexible Paxos is a variation of Mult-Paxos that allows different quorum sizes for first and second phases of the protocol as long as these two quorums intersect. Since the first phase is not executed as long as the leader is stable but the second phase is executed constantly, decreasing the number of processes to reach to a second quorum (also increasing the number of processes to reach to a first quorum), can increase the throughput by being capable of handling more failures.

\subsection{Linearization Points in Multi-Paxos}

We instrument Multi-Paxos with linearization points of successful QTree invocations. Multi-Paxos is a refinement of a set of QTree instances, one instance for each sequence number. The linearization points will refer to a specific instance identified using a sequence number, e.g., $sn.\add(r,v,r')$ denotes an $\add(r,v,r')$ invocation on the QTree instance $sn$. Therefore
\begin{itemize}
\item the linearization point of $sn.\add(r, v, r')$ occurs when the proposer broadcasts the {\fontfamily{qcr}\selectfont PROPOSE}($r$, $sn$) message containing value $v$ (during the {\bf \fontfamily{qcr}\selectfont PROPOSE} action in round $r$). $v$ is the value of the {\fontfamily{qcr}\selectfont JOIN}($r$, $sn$) message selected by the proposer. If $r' = 0$ then, $v$ is selected randomly.
\item the linearization point of $sn.\commit(r)$ occurs when the proposer who is the leader of the round $r$ updates $decidedVal$ for the sequence number $sn$.
\end{itemize}

\subsection{Multi-Paxos Refines A Set of QTree Instances}
\begin{theorem}
Multi-Paxos refines a composition of independent QTree instances.
\end{theorem}
\begin{proof}
We show that the sequence of successful $\add$ and $\commit$ invocations on a QTree instance $sn$ defined by linearization points along a Multi-Paxos execution satisfies the properties in Theorem~\ref{th:main} and therefore, it represents a correct QTree execution:
\begin{itemize}
\item \textbf{Property~\ref{eq:prop1}:} By definition, proposers can not propose two different proposals in the same round for the same sequence number. Since a leader can not propose for the next sequence number before deciding on the current one and round numbers are monotonically increasing when the leader is changed, the linearization point of $sn.\add(r,\_,\_)$ will occur at most once for a round $r$ with the same sequence number $sn$. Therefore, processes can only vote for a single propose with the same round and sequence number. This implies that at most one linearization point of $sn.\commit(r)$ can occur for a round $r$ with the same sequence number $sn$.
\item \textbf{Property~\ref{eq:prop0}:} This holds trivially as all the processes follow the rules of the protocol and they need to receive a {\fontfamily{qcr}\selectfont PROPOSE}($r$, $sn$) message (which can occur only after the linearization point of $sn.\add(r,\_,\_)$) from the leader of the current round to send {\fontfamily{qcr}\selectfont VOTE}($r$, $sn$) message.
\item \textbf{Property~\ref{eq:prop2}:}  In Multi-Paxos, leaders select the value that will be proposed for each sequence number separately, it can be accepted as running different instances {\bf \fontfamily{qcr}\selectfont PROPOSE} action of Single-decree Paxos. Therefore, the proof will follow as the proof of Property~\ref{eq:prop2} in Section~\ref{sec:history_events}, by considering that the property holds for each sequence number $sn$. Note that, skipping first phases in a round for the next sequence numbers after the first decision under a stable leader does not affect the proof because at the beginning of this round, the leader has already received highest votes (if there exist) from a quorum of processes for all sequence numbers. 
	\begin{itemize}
	\item \textbf{Property~\ref{eq:prop2a}:} It holds by the proof of Property~\ref{eq:prop2a} in Section~\ref{sec:history_events}, by considering it for each $sn$.
	\end{itemize}
\item \textbf{Property~\ref{eq:prop3}:}  Assume by contradiction that $sn.\commit(r)$ occurred along with the other two linearization points of $\add$. The linearization point of $sn.\commit(r)$ occurs because of a quorum of {\fontfamily{qcr}\selectfont VOTE}($r$, $sn$) messages sent by a set of processes $P_1$, and  $sn.\add(r',\_,r'')$ because of a quorum of {\fontfamily{qcr}\selectfont JOIN}($r'$) messages sent by a set of processes $P_2$. Since $P_1$ and $P_2$ must have a non-empty intersection, by the definition of the {\bf {\fontfamily{qcr}\selectfont JOIN}} action, it must be the case that $r'' \geq r$, which contradicts the hypothesis. 
\end{itemize}
\end{proof}

Cheap Paxos and Stoppable Paxos are just restricted versions of Multi-Paxos in which indices of sequences are independent. Therefore Multi-Paxos and both protocols refine QTree. Since the quorums are independent from which processes are included, Multi-Paxos will work the same with Cheap Paxos using the same processes. As a side note, in Cheap Paxos, since quorum of active (not idle) processes and any quorum decided a value must intersect, these values will be propagated the next rounds without any problem. In Stoppable Paxos, a decided value for a sequence number cannot turn into undecided or change its value due to a decided $stp$ value in another sequence number. Decided $stp$ value can only prevent execution on a higher sequence number. Since both Multi-Paxos and Stoppable Paxos progress the same until a $stp$ command is executed, Multi-Paxos and therefore Stoppable Paxos refine a set of QTree instances.

In Fast Paxos, when the processes receive request directly from the clients and propose accordingly, there can be multiple proposals due to network which are not proposed by the leader. Therefore we need to redefine the linearization points for Fast Paxos:
\begin{itemize}
\item the linearization point of $sn.\add(r,v,r')$ occurs when the leader broadcasts the {\fontfamily{qcr}\selectfont PROPOSE}($r$, $sn$) message containing value $v$ (during the {\bf \fontfamily{qcr}\selectfont PROPOSE} action in round $r$). $v$ is the value of the {\fontfamily{qcr}\selectfont JOIN}($r$, $sn$) message selected by the proposer. If $r' = 0$ then $v$ is selected randomly. If the proposer is not the leader, then $sn.\add(r,v,r')$ occurs when the leader of the round $r$ updates $decidedVal$ with value $v$ fo sequence number $sn$, after obtaining a quorum of votes for $v$. Note that $r' = 0$ and $v$ is selected randomly as no highest vote seen by the leader, for the sequence number $sn$.
\item the linearization point of $sn.\commit(r)$ occurs when the proposer who is the leader of the round $r$ updates $decidedVal$ with value $v$ fo sequence number $sn$.
\end{itemize}

Simply, when the processes receive the proposal from the client directly, the linearization point of $sn.\add(r,v,r')$ occurs at the same time with $sn.\commit(r)$ (if it is added) according to the definition of $sn.\commit(r)$. This is also intuitive and shows how this protocol received his name. Properties still hold because it is ensured that there was no vote beforehand for current sequence number if the request is directly received from the client and votes from these rounds are not considered when a new round starts:
\begin{itemize}
\item \textbf{Property~\ref{eq:prop1}:} Property~\ref{eq:prop1} in Multi-Paxos holds. Additionally, when the proposal is directly received from the client, as there can be only 1 quorum of fast votes for this proposal in the same round, there can be at most one linearization point of $sn.\add(r,\_,\_)$  and at most one linearization point of $sn.\commit(r)$.
\item \textbf{Property~\ref{eq:prop0}:} It holds by the proof of Property~\ref{eq:prop0} in Multi-Paxos.  
\item \textbf{Property~\ref{eq:prop2}:}  It holds by the proof of Property~\ref{eq:prop2} in Multi-Paxos. When the linearization point of $sn.\add(r,\_,\_)$ occurs after the proposal that is received by a process directly from the client, since $r' = 0$, this property is not considered.
 	\begin{itemize}
	\item \textbf{Property~\ref{eq:prop2a}:} It holds by the proof of Property~\ref{eq:prop2a} in Multi-Paxos. Again, when the linearization point of $sn.\add(r,\_,\_)$ occurs after the proposal that is received by a process directly from the client, since $r' = 0$, this property is not considered.
	\end{itemize}
\item \textbf{Property~\ref{eq:prop3}:} Assume by contradiction that $sn.\commit(r)$ occurred along with the other two linearization points of $\add$. The linearization point of $sn.\commit(r)$ occurs because of a quorum of {\fontfamily{qcr}\selectfont VOTE}($r$) messages sent by a set of processes $P_1$, and $sn.\add(r',\_,r'')$ exists because: 
\begin{itemize}
\item	There was only one value in highest votes for $sn$ (which is voted in round $r'$) in {\fontfamily{qcr}\selectfont JOIN}($r'$) messages sent by a set of processes $P_2$ or
\item	There were multiple values in highest votes (which are voted in round $r'$) but one of them is voted by at least $f+1$ of the processes in $P_2$.
\end{itemize}
In both cases, since $P_1$ and $P_2$ must have $f+1$ processes intersecting, by the definition of the {\bf {\fontfamily{qcr}\selectfont JOIN}} action, it must be the case that $r''\geq r$, which contradicts the hypothesis. 
\end{itemize}

Flexible Paxos refines QTree as the proof of properties perfectly fit for this protocol, without a modification on linearization points. In proof of properties for Multi-Paxos, the only two quorum that we are interested on their intersection is quorums from first and second phases (quorum of {\fontfamily{qcr}\selectfont JOIN} and  {\fontfamily{qcr}\selectfont VOTE} respectively, used in Property~\ref{eq:prop3}). Therefore, changing their sizes as long as they intersect doesn't affect the proof and the proof for Multi-Paxos holds as it is for Flexible Paxos.

\end{document}